\documentclass[11pt,a4paper]{article}
\usepackage{jcappub}

\usepackage{dcolumn}
\usepackage{bm}
\usepackage{xcolor}
\usepackage{float}
\usepackage{graphicx}
\usepackage{parskip}
\usepackage{mathtools}
\usepackage{subcaption}

\usepackage[english]{babel}

\def\beq{\begin{equation}}
\def\eeq{\end{equation}}
\def\beqa{\begin{eqnarray}}
\def\eeqa{\end{eqnarray}}

\newcommand{\p}{\vec{p}}
\renewcommand{\k}{\vec{k}}
\newcommand{\q}{\vec{q}}

\renewcommand{\r}{\vec{r}}
\newcommand{\x}{\vec{x}}
\renewcommand{\v}{\vec{v}}

\newcommand{\nl}{\nonumber \\}

\DeclareMathOperator\erf{erf}

\begin{document}
%\makeatletter

\title{Anatomy of astrophysical echoes from axion dark matter}

\author[a,b]{Elisa Todarello,} 
%\emailAdd{elisamaria.todarello@unito.it}
\author[c]{Francesca Calore} 
\author[a,b]{and Marco Regis} 
\affiliation[a]{Dipartimento di Fisica, Universit\`a di Torino, Via P. Giuria 1, 10125 Torino, Italy}
\affiliation[b]{Istituto Nazionale di Fisica Nucleare, Sezione di Torino, Via P. Giuria 1, 10125 Torino, Italy}
\affiliation[c]{LAPTh, CNRS, F-74000 Annecy, France}

\emailAdd{elisamaria.todarello@unito.it}
\emailAdd{francesca.calore@lapth.cnrs.fr}
\emailAdd{marco.regis@unito.it}

\abstract{
If the dark matter in the Universe is made of $\mu$eV axion-like particles (ALPs), then a rich phenomenology can emerge in connection to their stimulated decay into two photons. We discuss the ALP stimulated decay induced by electromagnetic radiation from Galactic radio sources. Three signatures, made by two echoes and one collinear emission, are associated with the decay, and can be simultaneously detected, offering a unique opportunity for a clear ALP identification. We derive the formalism associated with such signatures starting from first principles, and providing the relevant equations to be applied to study the ALP phenomenology.
We then focus on the case of Galactic pulsars as stimulating sources and derive forecasts for future observations, which will be complementary to helioscopes and haloscopes results.
}
\maketitle

%%%%%%%%%%%%%%%%%%%%%%%%%%%%%%%%%%%%
%%%%%%%%%%%%%%%%%%%%%%%%%%%%%%%%%%%%
\section{Introduction}
QCD axions are hypothetical particles introduced to solve the so-called strong CP problem~\cite{Peccei:1977hh,Peccei:1977ur, Weinberg:1977ma, Wilczek:1977pj}, namely to explain the extremely small value of the 
neutron's electric dipole moment. 
This pseudo-scalar Nambu-Goldstone boson, associated with the Peccei-Quinn symmetry breaking, can be produced in the early phase of the Universe, before or after inflation~\cite{Preskill:1982cy,Abbott:1982af,Dine:1982ah}.
In this way, the axion can also be a good candidate for cold dark matter (DM), for values of the Peccei-Quinn scale, $f_a$, of the order of $10^{10} - 10^{12}$ GeV. Such values of $f_a$, in turn, correspond to an axion mass, $m_a$, in the rage $10^{-6} - 10^{-3}$~eV. 
QCD axion represents a top-priority target for DM, and many detection strategies in the laboratory and in the sky have been designed in order to find it~\cite{Chou:2022luk}, exploiting its very weakly couplings with Standard Model (SM) particles and, in particular, with photons.

It was later realized that axion-like particles (ALPs) are also present in many  
beyond-the-standard-model theories, such as string 
theory models~\cite{Arvanitaki:2009fg}.
These ``cousins" of the QCD axion can be (pseudo-)scalar particles, with masses as low as zeV, 
with very weak couplings to the SM. As QCD axions do, they couple to photons through the Lagrangian
term $\mathcal{L} \propto g_{a\gamma} a F_{\mu \nu}\tilde{F}^{\mu \nu}$.
Moreover, they can also be good DM candidates in some portions of the parameter space. 

Thanks to their coupling with photons, besides the well-known possibility for an ALP to convert into a photon in the presence of external magnetic fields~\cite{Sikivie:1983ip}, an ALP has a finite probability to produce two photons from its decay.
Moreover, in the presence of an ambient radiation field, a stimulated enhancement of the decay rate occurs. In the radio band and for certain astrophysical environments, this can amplify the photon flux by several orders of magnitude.
The observational prospects concerning the ALP stimulated decay inside a photon emitting source, such as dwarf spheroidal galaxies, the Galactic Center, and galaxy clusters, were studied in \cite{Caputo:2018ljp,Caputo:2018vmy}.
The two photons arising from the decay are emitted back-to-back, i.e.~one with the same direction as the stimulating photon and one in the opposite direction (or, more precisely, approximately in the opposite direction since the ALP velocity is non-relativistic, but not zero).
This means that stimulated ALP decay generates ``echoes", i.e.~photons traveling in the opposite direction with respect to the flux of the stimulating source.
The production and detection of the ALP echo from an artificial beam was proposed in~\cite{Arza:2019nta} with 
in-depth calculation in~\cite{2022PhRvD.105b3023A} and a possible concrete experiment discussed in \cite{2023arXiv230906857A} (see also \cite{2023PhRvD.108h3001A, Gong:2023ilg}).
It was soon realized that also ``natural" beams, from the emission of astrophysical sources, could induce echoes in the radio sky~\cite{2020arXiv200802729G}.
In particular, the potential of the echo signatures in our Galaxy was explored for supernova remnants, which are particularly promising since they could have been significantly brighter in the past~\cite{2022PhRvD.105g5006B,2022PhRvD.105f3007S}.

As already noted in early works~\cite{Preskill:1982cy, Abbott:1982af,
Dine:1982ah, Tkachev:1986tr, Tkachev:1987cd}, if the coherence length of the axion field is larger than the typical distance a photon can travel before stimulating the decay of an axion, the photon field can grow exponentially due to parametric resonance~\cite{ Kephart:1994uy,Masso:1997ru,Yoshida:2017ehj, Arza:2018dcy, McDonald:2019wou, Arza:2019nta,
Sigl:2019pmj, Wang:2020zur, Chen:2020ufn,Lee:1999ae,Alonso-Alvarez:2019ssa,Tkachev:2014dpa, Hertzberg:2018zte,Rosa:2017ury,Ikeda:2019fvj, Arza:2020eik,Levkov:2020txo, Du:2023jxh, Chung-Jukko:2023cow}.

The main aim of this work is to present a full and complete derivation of the astrophysical echoes from ALP DM, starting from first principles, and determining the dependence of the signal on the DM properties (such as the spatial and velocity distributions) and the source characteristics (such as distance, age, and motion). 
We will discuss three signatures, made by two echoes and one collinear emission, associated with the decay, which can be simultaneously detected, offering a unique opportunity for a clear ALP identification.

The paper is organized as follows.
In Section~\ref{sec:echo_signal}, we describe the derivation of the echo signal from a point source, with additional details of the computation reported in Appendices~\ref{sec:signal}-\ref{sec:implementation}. Section~\ref{sec:sky} describes the observational properties of the echoes and their dependence on the DM halo and source characteristics. In Section~\ref{sec:collem} and Appendix~\ref{sec:collin}, we treat the case of collinear emission. Then we focus on the case of Galactic pulsars as stimulating sources, deriving observational prospects in Section~\ref{sec:forec}.
Section~\ref{sec:conc} concludes.

%%%%%%%%%%%%%%%%%%%%%%%%%%%%%%%%%%%%
%%%%%%%%%%%%%%%%%%%%%%%%%%%%%%%%%%%%
\section{Echo signal from a point source}\label{sec:echo_signal}
\begin{figure}[t]
\centering
\includegraphics[width=\linewidth]{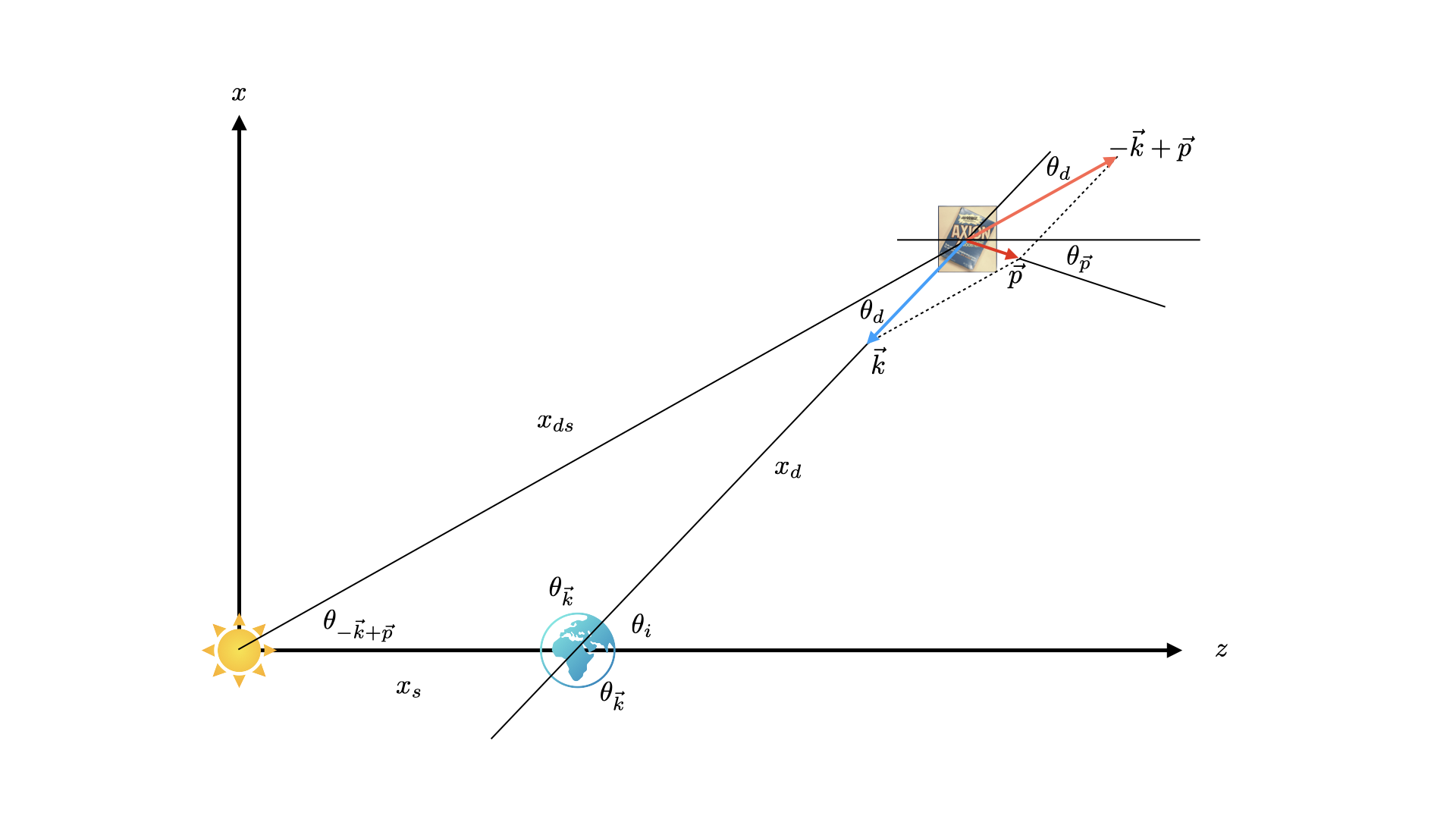}
\caption{Geometry of the back-light echo.}
\label{fig:geom1}
\end{figure}
\begin{figure}
  \centering
\includegraphics[width=\linewidth]{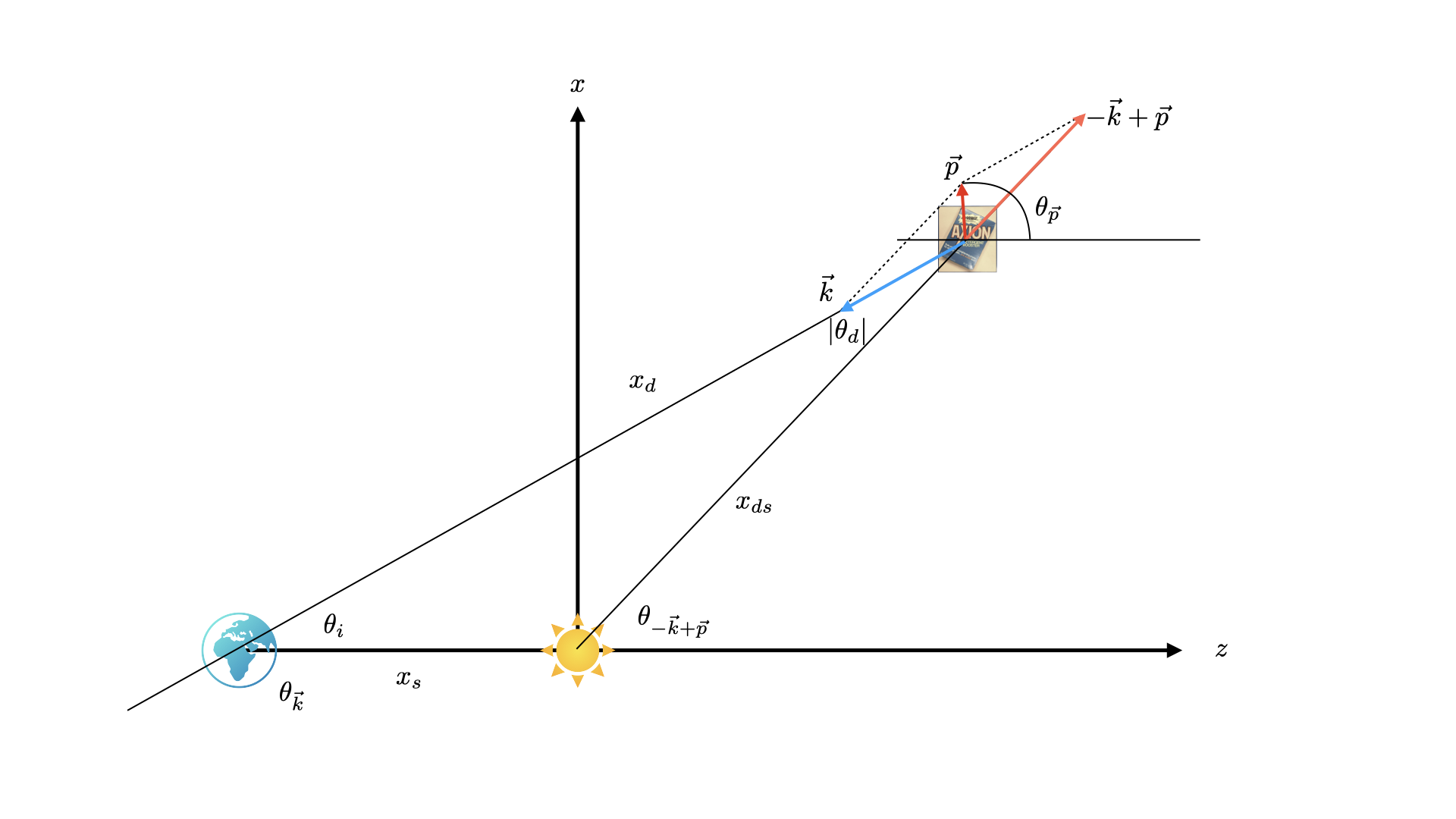}
\caption{Geometry of the front-light echo.}
\label{fig:geom2}
\end{figure}

We want to describe the following process: a photon emitted by a source with momentum $\q$ stimulates the decay of a DM ALP\footnote{In the following, we use the terms axion and ALP interchangeably.} with momentum $\p$. Two photons are produced in the decay: one with momentum $\q$, identical to the incoming photon, and one with momentum $\k$, the echo photon.
In this Section, our task is to derive an expression for the flux density of the echo as seen from Earth. We will treat the collinear emission in Section~\ref{sec:collem}. 

The Boltzmann equation for the axion number density $n_a$ tells us how many axions decays occur per unit time:
\beqa
\frac{d^3\dot{n}_a}{dk^3}
 &=& -
\frac{1}{(2\pi)^3}\frac{1}{2\omega_k}
 \int \frac{d^3p}{(2\pi)^3}\frac{f_a(\p\,)}{2m_a}\ 
 \frac{d^3q}{(2\pi)^3}\frac{1}{2\omega_q}
 (2\pi)^4\delta^{(4)}(q + k - p) \sum_{\lambda}|\mathcal{M}_0|^2
 (1+f_\gamma(\q, \lambda)) 
 \enspace.\nl\label{boltzmann}
\eeqa
Here $ \sum_{\lambda}|\mathcal{M}_0|^2$ is the matrix element squared, summed over the polarizations of the two final state photons, and  $f_a$ and $f_\gamma$ are the phase space distributions of the axions and of the photons emitted by the source.
Notice that the enhancement factor from phase space density of photons with momentum $\k$ does not appear in Eq.~\eqref{boltzmann} because we are considering a configuration in which this state is not initially highly occupied.

Let's focus on a point source located at the origin of our coordinate system emitting radially and isotropically. We borrow part of our notation from Ref.~\cite{2022PhRvD.105f3007S}. We consider an ALP decay happening at a location $\x_{ds}$ that we choose to lie on the $xz$-plane. At $\x_{ds}$, the photons from the source have a unique direction
\beqa
f_\gamma(\q, \x_{ds}) &=& f_\gamma(\omega_q, x_{ds})
\frac{\delta(\theta_{\q} - \theta_s)\delta(\phi_{\q})}{\sin\theta_{\q}} \enspace,\label{fgamma}
\eeqa
with $\theta_s$ determined by $\x_{ds}$.
% , for example as
% \beqa
% \cos\theta_s = \frac{z_{ds}}{r_{ds}}\enspace.
% \eeqa
We take the Earth to lie on the $z$-axis. The geometry is shown in Figures~\ref{fig:geom1} and \ref{fig:geom2}.
We consider two possible configurations: 1) the \textit{back-light echo} coming from the direction opposite to the source, shown in Figure~\ref{fig:geom1} and 2) the \textit{front-light echo} coming from behind the source, shown in Figure~\ref{fig:geom2}. 

For the axion phase space distribution, we choose a Maxwell-Boltzmann
\beq
f_a(\p,\x\,) = n_a \frac{(2\pi)^3}{(2\pi)^{3/2}\delta p(\x\,)^3}\ e^{-\frac{(\p - \langle\p\,\rangle)^2}{2\delta p(\x\,)^2}}\enspace,\label{fa}
\eeq
where $\langle\p\,\rangle$ is the average DM momentum in Earth's rest frame and $\sqrt{3}\delta p(\x\,)$ is the total momentum dispersion at a location $\x$ assuming no anisotropy.
The axion number density is given by
\beq
n_a(\x\,) = \int\frac{d^3p}{(2\pi)^3}\, f_a(\p\,,\x\,)\enspace. \label{na}
\eeq
From Eq.~\eqref{boltzmann}, with Eqs.~\eqref{fgamma}
and~\eqref{fa}, through the derivation described in  Appendix~\ref{sec:signal}, we obtain 
\beqa
\frac{d\dot{n}_a}{d\omega_k d\Omega_k}
&=& -  n_a \,
\frac{g^2 m_a^2}{256\pi^2} \frac{N_{pol}}{2}\, f_\gamma(\omega_k)\ h(\omega_k, \x_{ds})\enspace,\label{d3ndk3}
\eeqa
with
\beqa
h(\omega_k, \x_{ds})
= \frac{1}{(2\pi)^{3/2}\delta v^3}
\ e^{-\frac{\langle v_t\rangle^2}{2\delta v^2}}
\ e^{-\frac{(\varepsilon - \langle v_\parallel\rangle)^2}{2\delta v^2}}
\ e^{-\frac{(\omega_k\theta_d/m_a - \langle v_\perp\rangle)^2}{2\delta v^2}} \label{h}
 \enspace,
\eeqa
where we have used the fact that for non-relativistic axions $\p=m_a\v$.
In the equations above, $\Omega_k$ is the direction pointing from the decay location to the detector, $\langle v_\parallel\rangle$ and $\langle v_\perp\rangle$ are the components of the axion average velocity parallel and perpendicular to the line of sight (los) in the $xz$-plane, $\langle v_t\rangle$ is the average DM velocity perpendicular to the $xz$-plane, $\varepsilon = 2\omega_k/m_a - 1$, and $\theta_d$ is the angle between the los and the direction from the source to the decay location.
In Eq.~\eqref{d3ndk3}, $n_a$, $f_\gamma$, $\langle \v\,\rangle$ and $\theta_d$ are all functions of the decay location $\x_{ds}$. Explicit expressions for the components of $\langle \v\,\rangle$ are given in Appendix~\ref{sec:implementation}. 

The interpretation of Eq.~\eqref{d3ndk3} is the following. At the decay location $\x_{ds}$, the source's photons arrive from a certain direction $\theta_s$. Any ALP present at $\x_{ds}$ can be stimulated to decay, but only if the ALP in question has velocity components $v_\parallel = \epsilon$,  $v_\perp = \omega_k\theta_d/m_a$ and $v_t=0$ will the echo photon travel in the direction of Earth with frequency $\omega_k$.
The likelihood of such a decay depends then on how many axion particles are available with the required momentum.

For a fixed observation angle $\theta_i \ll 1$, the deflection angle $\theta_d$ takes the following form
\beq
\theta_d = \pm \mathrm{arcsin}\left(\frac{x_s}{x_{ds}}\theta_i\right)\enspace,
\eeq
where the upper sign corresponds to the back-light echo, the lower to the front-light echo.
For the back-light echo, $x_s < x_{ds}$, so that a small $\theta_i$ implies a small $\theta_d$.
For the front-light echo, instead,  $x_s/x_{ds}$ can be large if the decay happens close to the source. Thus, in this case, a small $\theta_i$ does not necessarily imply a small $\theta_d$. However, for non-relativistic axions, decays with large $\theta_d$ are suppressed by the axion phase space distribution. So, in both cases, we can approximate
\beq
\theta_d = \pm \frac{x_s}{x_{ds}}\theta_i\enspace.
\eeq

The decay rate is the largest at a frequency 
\beq
\omega_* 
\approx \frac{m_a}{2}\left( 1 + \langle v_\parallel\rangle \right)
\enspace,
\eeq
and is suppressed exponentially for $\omega - \omega_* > m_a\delta v$.
Since $\langle \v\,\rangle$ is small, we can approximate $\omega_k \sim m_a/2$ in Eq.~\eqref{d3ndk3}, keeping corrections of order $\langle \v\,\rangle$ only in the exponents.

The rate of change of the axion number density can be related to the echo flux density observed from Earth.
For each axion decay, two photons, each of energy $\sim m_a/2$ are produced. Only one of those travels toward Earth. 
The power emitted towards us per unit volume, unit frequency, and unit solid angle is then equal to $m_a/2\ |d\dot{n}_a / (d\nu_k d\Omega_k)|$. 
Integrating along the los, we obtain the spectral intensity
\beq
I_\nu = \int dx_d \ \frac{m_a}{2} \Big|\frac{d\dot{n}_a}{d\nu_k d\Omega_k}\Big|\enspace,
\eeq
where we assumed absorption effects to be negligible.
Next, we need to relate the source's photon phase space density to its flux density observed from Earth. Using Eq.~\eqref{fgamma}, we obtain 
\beq
f_\gamma(\omega, x_{ds}) = \frac{(2\pi)^3}{N_{pol}\omega^3}\frac{d\rho_\gamma(x_{ds})}{d\omega}
= \frac{(2\pi)^3}{N_{pol}\omega^3}\frac{d\rho_\gamma(x_s)}{d\omega} \left(\frac{x_s}{x_{ds}} \right)^2\enspace,\label{fgamma_rhogamma}
\eeq
where $\rho_\gamma$ is the photon energy density. The flux density through a surface area perpendicular to the direction Earth-source is $S^{(0)}_\nu (x_s) = 2\pi d\rho_\gamma(x_s)/d\omega$.

Finally, from $S_\nu = \int d\Omega_i \,B(\Omega_i)\int_{\rm los} dI_\nu $ we obtain
\beq
S_\nu = \frac{\pi g^2}{16m_a}\ S_\nu^{(0)} 
\int d\Omega_i \,B(\Omega_i)\int dx_d\ \rho_a(\x_{ds})\, \left(\frac{x_s}{x_{ds}} \right)^2
h(\omega_k, \x_{ds})\enspace.\label{flux_density}
\eeq
where $S_\nu^{(0)}$ is the source's flux density as observed from Earth, $\Omega_i$ is the solid angle in the direction of the los $(\theta_i,~\phi_i)$, and $B$ describe the observing beam.

In Eq.~\eqref{flux_density}, we have not taken into account two possible effects: the motion of the source with respect to Earth and the time dependence of the source's luminosity.
If we introduce these effects, Eq.~\eqref{flux_density} generalizes to
\beq
S_\nu(t) = \frac{\pi g^2}{16m_a}\  S^{(0)}_\nu(t) 
\int d\Omega_i \,B(\Omega_i)\int dx_d\ \rho_a(\x_{ds})\,
\frac{L_\nu(t_{em})}{L_\nu(t - \tilde x_s(t))}
\left(\frac{\tilde x_s(t)}{\tilde x_{ds}(t_{em})}\right)^2 
h(\omega, \x_{ds}, t_{em})\enspace,\label{flux_density_gen}
\eeq
where $\tilde x_s(t)$ is the distance from Earth to the actual position of the source at time $t$ (not to the position where the source appears to be), $\tilde x_{ds}(t)$ is the distance from the source to the decay location at time $t$, and $L_\nu$ is the luminosity of the source.
At time $t_{em} = t - x_d - \tilde x_{ds}(t_{em})$, the source emitted a photon that subsequently traveled a distance $\tilde x_{ds}(t_{em})$ to the decay location, stimulated the decay of an ALP which in turn created a photon that traveled a distance $x_d$ to Earth, arriving at a time $t$. The time $t_{em}$ can be expressed as a function of $x_d$ and the velocity of the source $\v_s$ as described in Appendix~\ref{sec:motion}. The function $h$ is evaluated at a fixed location, determined by $\theta_i$ and $x_d$ independent of $t_{em}$. However, it develops a time dependence through $\theta_d$. This will become clear in Section~\ref{sec:source}.

%%%%%%%%%%%%%%%%%%%%%%%%%%%%%%%%%%%%
%%%%%%%%%%%%%%%%%%%%%%%%%%%%%%%%%%%%
\section{Angular distribution of the echo on the sky}\label{sec:sky}

To study the angular distribution of the signal in the sky, we isolate from $S_\nu$ the following dimensionless quantity
\beqa
g(\theta_i, \phi_i)  
&=& x_s\int_{r_{min}}^R dx_d\ 
\frac{\rho_a(\x_{ds})}{\rho_\odot}\ 
\frac{L_\nu(t_{em})}{L_\nu(t - x_s)}\ 
\frac{1}{x_{ds}(t_{em})^2}\ 
(2\pi)^{3/2}\delta v_\odot^3\ h(\omega, \x_{ds}(t_{em}))\enspace,
\label{g}
\eeqa
where $\rho_\odot$ and $\delta v_\odot$ are the DM density and velocity dispersion at the Sun's location.

\subsection{Idealized case}
\begin{figure}[t]
  \centering
  \includegraphics[width=\textwidth]{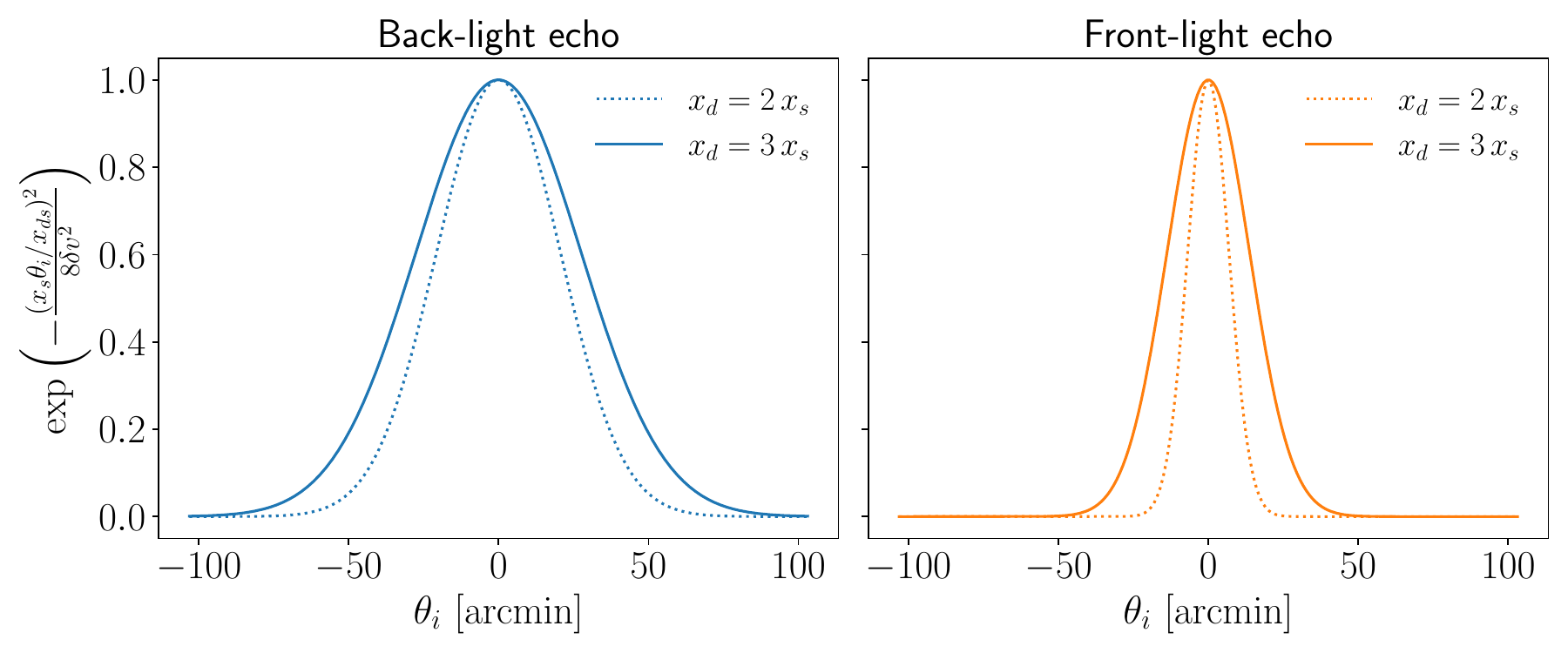}
\caption{Angular distribution of the echo from decays happening at fixed distance $x_d$ in the idealized case. We set $\delta v = 10^{-3}$. }
\label{fig:sky_at_xd}
\end{figure}

As a first step, we want to establish a basis onto which to add the various effects relevant to the distribution on the sky of the echo intensity.
To this end, we consider the academic case of an infinite DM halo with constant energy density
$\rho_a(\x_{ds}) = \rho_\odot$, velocity dispersion $\delta v(\x_{ds}) = \delta v_\odot$, and null DM and source velocity $\langle\v\,\rangle = 0$. We also assume constant source luminosity $L_\nu(t)$ and velocity $\v_s=0$.
We fix the energy such that we are at resonance $\omega = \omega_*$.
Under these assumptions, $g(\theta_i, \phi_i)$ reduces to
\beqa
g_{id}(\theta_i)  
&=& x_s\int_{r_{min}}^R dx_d\ 
\frac{1}{x_{ds}^2}\ 
\ \exp\left(-\frac{\left(\frac{x_s}{x_{ds}}\theta_i\right)^2}{8\delta v^2}\right)\enspace.
\eeqa

A first observation is that, if in addition to requiring that the DM is at rest on average, we also require it to be perfectly cold, we obtain $g_{id} \propto \delta(\theta_i)$. This is to be expected because, in the absence of average velocity and velocity dispersion, the echo travels back to the location where the source was when the photons that stimulated the decay were emitted. For a source at rest emitting radially, the only way to receive the echo radiation is to be between the source and the decay location.

When velocity dispersion is present, the intensity on a spherical shell of radius $x_d$ centered at Earth drops exponentially as $\theta_i$ increases beyond $\frac{x_s}{x_{ds}}\theta_i \gtrsim 2\delta v$.  For fixed $x_d$, we thus have the typical size of the intensity patch
\beq
\theta_{i,0} \sim 2\delta v\left(\frac{x_d}{x_s} \pm 1\right)\enspace,
\eeq
where we have approximated $x_{ds}=x_d\pm x_s$. Again, the upper sign holds for the back-light echo, the lower sign for the front-light echo.
For given $x_d$, the intensity of the front-light echo is more compact than that of the back-light echo, because the distance to the source is smaller. We also notice that, for decays happening on shells of increasing $x_d$, the size of the patch becomes larger, as shown in Figure~\ref{fig:sky_at_xd}.
Correspondingly, for fixed $\theta_i$, the back-light echo emission is peaked at $x_d = 0$ if $\theta_i \lesssim 2\delta v$, while for larger observation angles $\theta_i > 2\sqrt{2}\delta v$, the decay needs to happen at larger distances in order to have the same $\theta_d$. In this case,
the largest contribution is from decays happening at $x_d = x_s (\theta_i/(2\sqrt{2}\delta v) - 1)$ and the flux is suppressed by the distance.
For the front-light echo, the flux from a given direction $\theta_i$ is always suppressed at distances $x_d < x_s (\theta_i/(2\sqrt{2}\delta v) + 1)$.
In both cases, the integrand decreases as $x_d^{-2}$ at large distances.

\begin{figure}[t]
  \centering
  \includegraphics[width=0.5\textwidth]{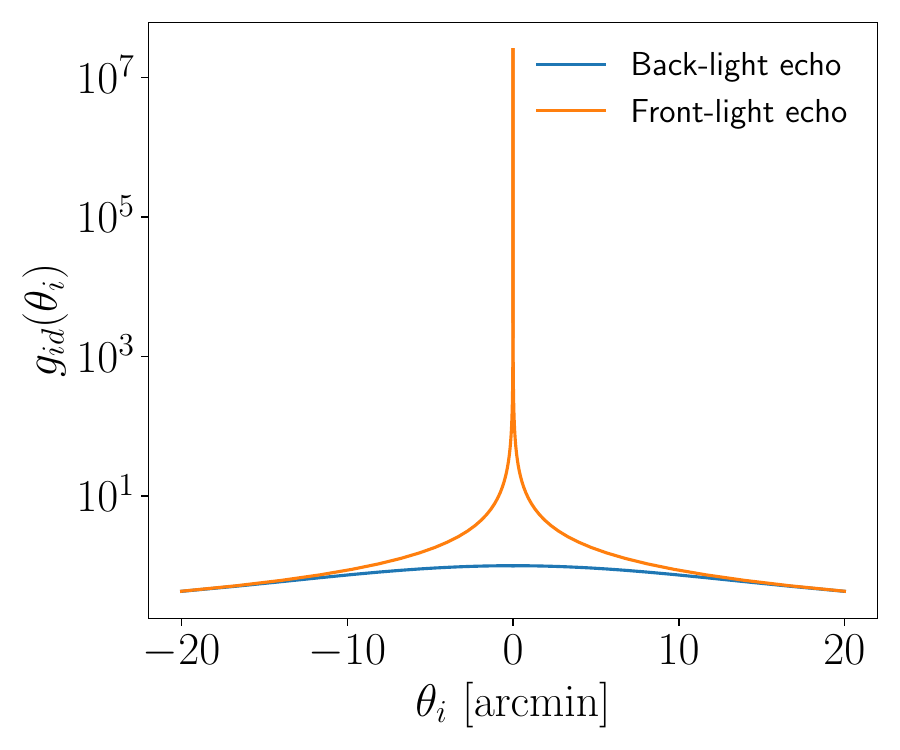}
\caption{Emission profile of the echoes after integrating along the los in the idealized case. We set $\delta v = 10^{-3}$.} 
\label{fig:sky_integrated}
\end{figure}

Carrying on with our idealized case, we can integrate over $x_d$ analytically in the approximation $x_{ds}\approx x_d\pm x_s$. 
The lower integration limit is Earth's radius for the back-light echo, while for the front-light echo we take $r_{min} = x_s$.
Sending $R\to\infty$, the result is
\beqa
g_{id}(\theta_i)
&=& \frac{\sqrt{2\pi}\delta v}{\theta_i}
\erf{\left(\frac{x_s\theta_i}{2\sqrt{2}\delta v (r_{min}\pm x_s)}\right)}\enspace.\label{gid_integrated}
\eeqa
For the back-light echo, we can neglect $r_{min}$ compared to $x_s$. Eq.~\eqref{gid_integrated} reduces to
\beqa
g_{id}^{(back)}(\theta_i)
&=& \frac{\sqrt{2\pi}\delta v}{\theta_i}
\erf{\left(\frac{\theta_i}{2\sqrt{2}\delta v }\right)}\enspace,\label{gid_integrated_oppo}
\eeqa
Notice that the result is finite for $\theta_i=0$
\beqa
g_{id}^{(back)}(0) &=& 1
\enspace.
\eeqa
For the front-light echo we can set the error function in Eq.~\eqref{gid_integrated} equal to 1, obtaining
\beqa
g_{id}^{(front)}(\theta_i) &=& \frac{\sqrt{2\pi}\delta v}{|\theta_i|}\label{gid_integrated_behi}
\enspace.
\eeqa
The divergence at $\theta_i=0$ is due to the assumption of a point source. The source's flux density diverges at the source location. In the case of the front-light echo, we can receive photons produced in decays happening very close to the source ($x_{ds}\approx 0$), which cause the divergence at $\theta_i = 0$, while this is not the case for the back-light echo. In practice, the finite size of the source will regulate the divergence.

%is only present in this hypothetical case, of an infinitely extended halo and an infinitely old source, since $R$ will be limited by the age of the source and the size of the Galaxy, once we introduce these effects.
% Moreover, the echo radiation from angles smaller than the radius of the source over $x_s$ is blocked by the source itself.

The idealized distribution on the sky is shown in Figure~\ref{fig:sky_integrated}.
The intensity is almost constant for $\theta_i \lesssim \delta v$ for the back-light echo, while for large $\theta_i$ it drops as $\theta_i^{-1}$. The typical size of the intensity patch on the sky is 10$^\prime$, taking $\delta v \sim 10^{-3}$. For the front-light echo, the intensity decreases as  $\theta_i^{-1}$ everywhere.

Note also that the echo emission can be disentangled from the direct radiation from the source in the front-light echo case by using polarization.
Indeed the echo polarization is orthogonal to the one of the source's emission.

%%%%%%%%%%%%%%%%%%%%%%%%%%%%%%%%%%%%
\subsection{Realistic case}\label{sec:realistic_case}
\begin{table}
    \centering
    \begin{tabular}{|c|c|c|c|c|c|c|}
        \hline
        Source name & $l~(^\circ)$ & $b~(^\circ)$ & $x_s~(\mathrm{kpc})$ & $\tau_{age}~(\mathrm{yrs})$ & $l_{\v_s}~(^\circ)$  & $b_{\v_s}~(^\circ)$\\
        \hline
         A & 90 & 0 & 0.1 & $5\times 10^3$ & 90 & 0\\
        \hline
         B & 270 & 45 & 1 & $10^4$ & 90 & 45\\
        \hline
         C & 0 & 90 & 0.5 & $5\times 10^4$ & 270 & 0\\
        \hline
         D & 0 & 0.1 & 3 & $10^5$ & 180 & 45\\
         \hline
    \end{tabular}
    \caption{Properties of hypothetical sources. The columns are the source Galactic longitude $l$, Galactic latitude $b$, distance from Earth as seen today $x_s$, age $\tau_{age}$ and direction of the source's velocity in Galactic coordinates $l_{\v_s}$ and $b_{\v_s}$. For all sources the magnitude of $\v_s$ is 200~km/s.}
    \label{tab:hypo_source}
\end{table}

Now that we have identified the main features of the distribution on the sky of the echo intensity, we proceed to turn on one by one the effects coming from the characteristics of the DM halo and of the source.

Whenever an effect is not taken into consideration, the corresponding quantity in Eq.~\eqref{g} is set to its default value. The default values are: $\rho(r)=\rho_\odot$, $x_s = 0.1$~kpc, $\delta v(r) =\delta v_\odot$, $\langle v_\perp\,\rangle = 0$, the age of the source $\tau_{age} = \infty$, $\v_s=0$ and $L=const$. In all plots of the next two subsections, we set the $\omega=\omega_*$ and we vary the observation angle $\theta_i$ leaving the Galactic longitude of the los direction unchanged. Thus, when we talk about negative (positive) $\theta_i$, we mean that we are observing directions of lower (larger) Galactic latitude than the source.

We illustrate the effects on four hypothetical sources with properties listed in Table~\ref{tab:hypo_source}.

%%%%%%%%%%%%%%%%%%%%%%%%%%%%%%%%%%%%
\subsubsection{Effects of the dark matter halo}\label{sec:halo}
\begin{figure}
  \centering
  \includegraphics[width=0.8\textwidth]{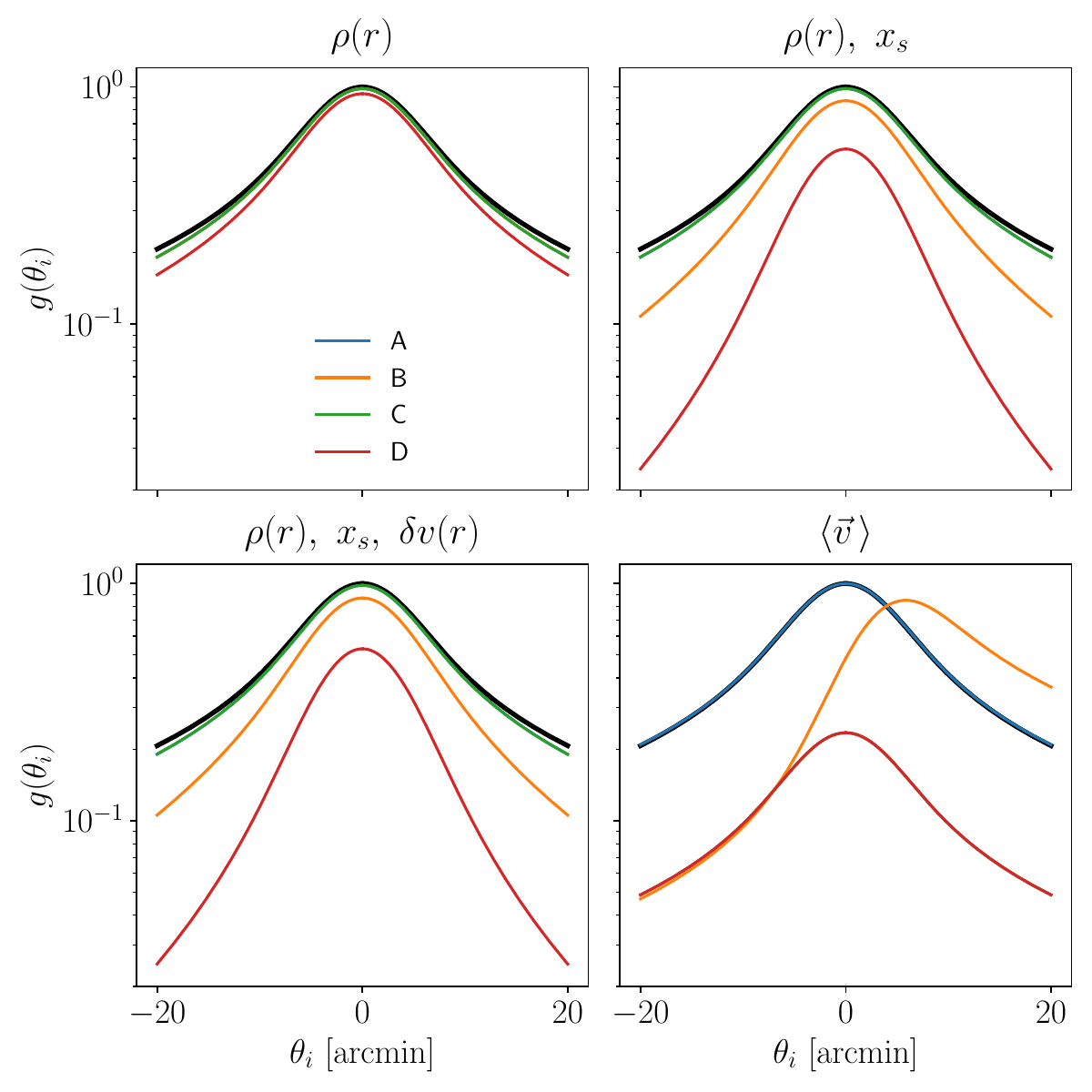}
\caption{Effects of the dark matter halo on the back-light echo.}
\label{fig:sky_integrated_ideal_opposite_halo}
\end{figure}
\begin{figure}
  \centering
  \includegraphics[width=0.8\textwidth]{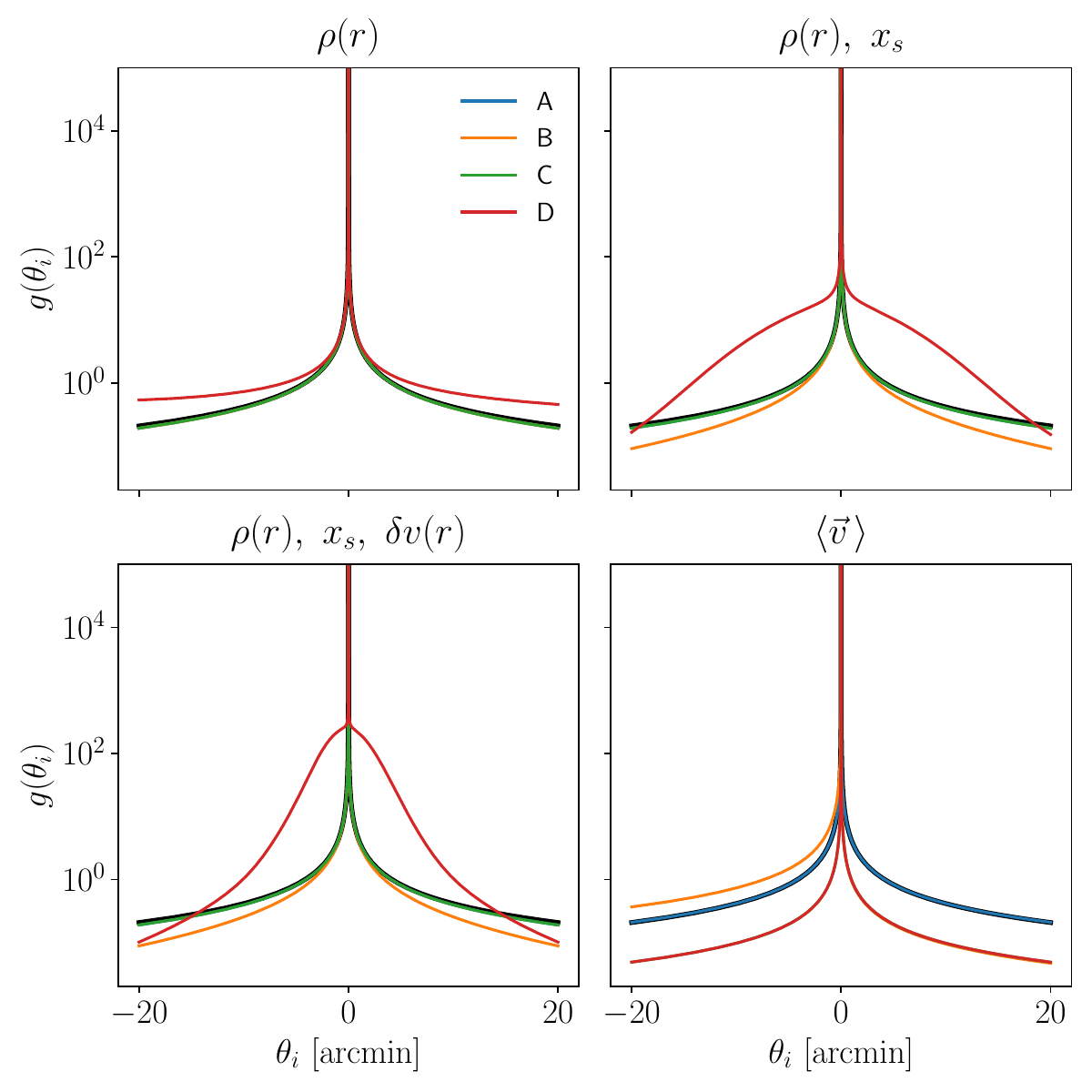}
\caption{Effects of the dark matter halo on the front-light echo.}
\label{fig:sky_integrated_ideal_behind_halo}
\end{figure}
We consider a Navarro-Frenk-White halo, with parameters $\rho_s = 0.41~\mathrm{GeV~cm}^{-3}$ and $r_s = 20$~kpc~\cite{Benito:2020lgu}, yielding a DM density at the Sun's location $\rho_\odot = 0.5~\mathrm{GeV~cm}^{-3}$. 
% where $\rho_\odot$ is the DM energy density at the Sun's galactocentric radius $r_\odot =  8.178$~kpc~\cite{GRAVITY}.
We calculate $\delta v(r)$ integrating the spherically symmetric Jeans equation assuming no velocity anisotropy (see for example~Ref.~\cite{Lokas:2000mu}). We obtain $\delta v_\odot=144$~km/s. 

We assume that the DM is on average at rest in a non-rotating frame attached to the Galaxy.
Then the average velocity of DM in Earth's rest frame $\langle \v\,\rangle$ is mainly given by 
$-\v_{\odot, \phi}$, where $\v_{\odot, \phi}$ is the velocity of the Sun in the tangential direction in the Galactic plane with magnitude 245.6~km/s~\cite{Drimmel_2018} in the direction of Galactic coordinates $(l, b) = (90^\circ, 0^\circ)$. We neglect the Sun's velocity in the other directions as well as the motion of Earth with respect to the Sun.

We study the effect on the echo of the following quantities:
\begin{enumerate}
    \item $\rho(r)$,
    \item $\rho(r)$ in combination with the distance of the source $x_s$,
    \item $\rho(r)$ in combination with $x_s$ and the velocity dispersion $\delta v(r)$,
    \item the DM velocity relative to the Sun $\langle\v\,\rangle$.
\end{enumerate}

In Figures~\ref{fig:sky_integrated_ideal_opposite_halo} and~\ref{fig:sky_integrated_ideal_behind_halo}, we show each of the effects above in a separate panel, for the back-light and front-light echo, respectively.
The thick black line in all panels represents the idealized case $g_{id}(\theta_i)$. 

On the top left panel of each figure, we see the effect of the DM density profile. The lines for sources A, B, and C are on top of each other because the DM density along the los is the same in these three cases. We observe a minor suppression of the echo flux compared to the idealized case. On the other hand, for source D, the back-light echo flux is more significantly reduced because the los points away from the Galactic center. As expected, we see an increased flux for source D in the case of the front-light echo.

The top right panels show the combined effect of the density profile and the distance to the source. Notice, that if we let $x_s$ vary among sources while keeping $\rho = \rho_\odot$ fixed, there would be no effect on the $g$ (see Eqs.~\eqref{gid_integrated_oppo},~\eqref{gid_integrated_behi}). The lines for sources A and C are unchanged from the top left panels. For sourced B and D, whose distances are larger than the default value, we see a reduction of the flux for the back-light echo due to the lower DM density along the los. Conversely, for the front-light echo, we notice a substantial increase in flux for source D which is now placed closer to the Galactic center. This is due to two facts: the DM density is higher where $x_{ds}\sim 0$, and the contribution from the high DM density at the Galactic center is not suppressed by a large value of $x_{ds}$.

The bottom left panels display the effect of the variation of the velocity dispersion along the los in combination with the source distance and the variation of the DM energy density.
In all cases, except for the front-light echo of source D, the effect is negligible. The velocity dispersion goes to zero at large Galactic radii, causing a reduction of the echo flux from these locations. However, in this case, the flux is already suppressed by the distance and by the low DM density. 
We see instead a non-negligible effect on the front-light echo of source D, for which the los passes close to the Galactic center, where the velocity dispersion is small.
For observation angles small enough to have $\theta_d < \delta v$ close to the Galactic center, we see an enhancement of the flux due to the normalization of $h(\omega, \x_{ds})$, i.e.~there are more axions available to decay that will produce photons with the desired direction.
If the observation angle is larger than that, the suppression from the exponential wins over the enhancement from the normalization factor and we observe a reduction of the flux.

The bottom right panel of each figure shows the effect of the average DM velocity. For source A, $\langle\v\,\rangle$ is almost completely along the los. Hence it does not affect the distribution of the echo on the sky, but rather it will shift its frequency. For source B, instead, we have a sizeable value of $\langle\v_\perp\rangle$.
According to Eq.\eqref{h}, the echo flux is the largest when $\omega_k \theta_d / m_a = \langle v_\perp \rangle$ or
\beq
\theta_i = \pm \frac{m \langle v_\perp \rangle}{\omega_k}\frac{x_{ds}}{x_s}
\eeq
For the back-light echo, the emission is peaked at a location where $\theta_i$ has the same sign as $\langle v_\perp \rangle$.  Now we have to be careful to interpret this correctly.
From Eq.~\eqref{pxpz}, we see that $\langle v_\perp \rangle$ has the opposite sign as $\langle v_x\rangle$ ($\cos\theta_{\k}<0$).
Since $\theta_i$ grows in the positive $x$-direction, the peak of the back-light echo emission comes from the direction opposite to that of $\langle v_x\rangle$. In other words,
the peak of the back-light echo emission is shifted in the direction opposite to that of the average DM velocity.
When we look in the direction opposite to source B, the component of the average DM velocity perpendicular to the los points in the direction of decreasing Galactic longitude, i.e. $\langle v_x\rangle < 0$, causing the emission to peak at positive $\theta_i$. The contrary is true for the front-light echo, for which the largest emission comes from the direction where the DM velocity points.
However, in this case, we don't see a shift of the peak, but only an asymmetric distribution of the flux.
For sources C and D, whose lines are on top of each other, we have $|\langle\v\,\rangle| \approx |\langle v_t\rangle|$, resulting in an overall reduction of the flux by a factor $\exp\left[-\langle v_t\rangle^2 / 2\delta v^2\right]$.

%%%%%%%%%%%%%%%%%%%%%%%%%%%%%%%%%%%%
\subsubsection{Effects of the source's characteristics}\label{sec:source}
\begin{figure}
  \centering
  \includegraphics[width=0.8\textwidth]{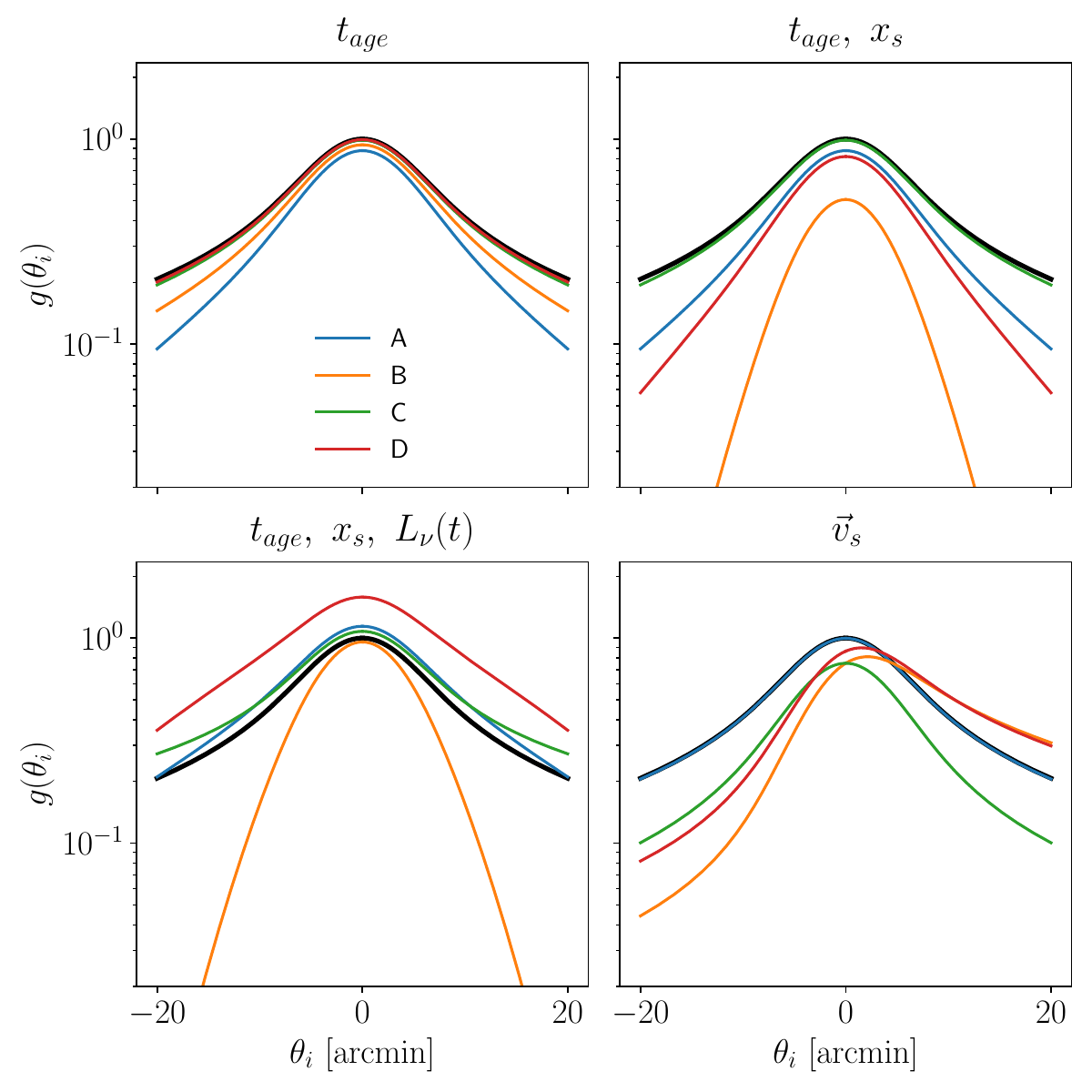}
\caption{Effects of the source characteristics on the back-light echo.}
\label{fig:sky_integrated_ideal_opposite_source}
\end{figure}
\begin{figure}
  \centering
  \includegraphics[width=0.8\textwidth]{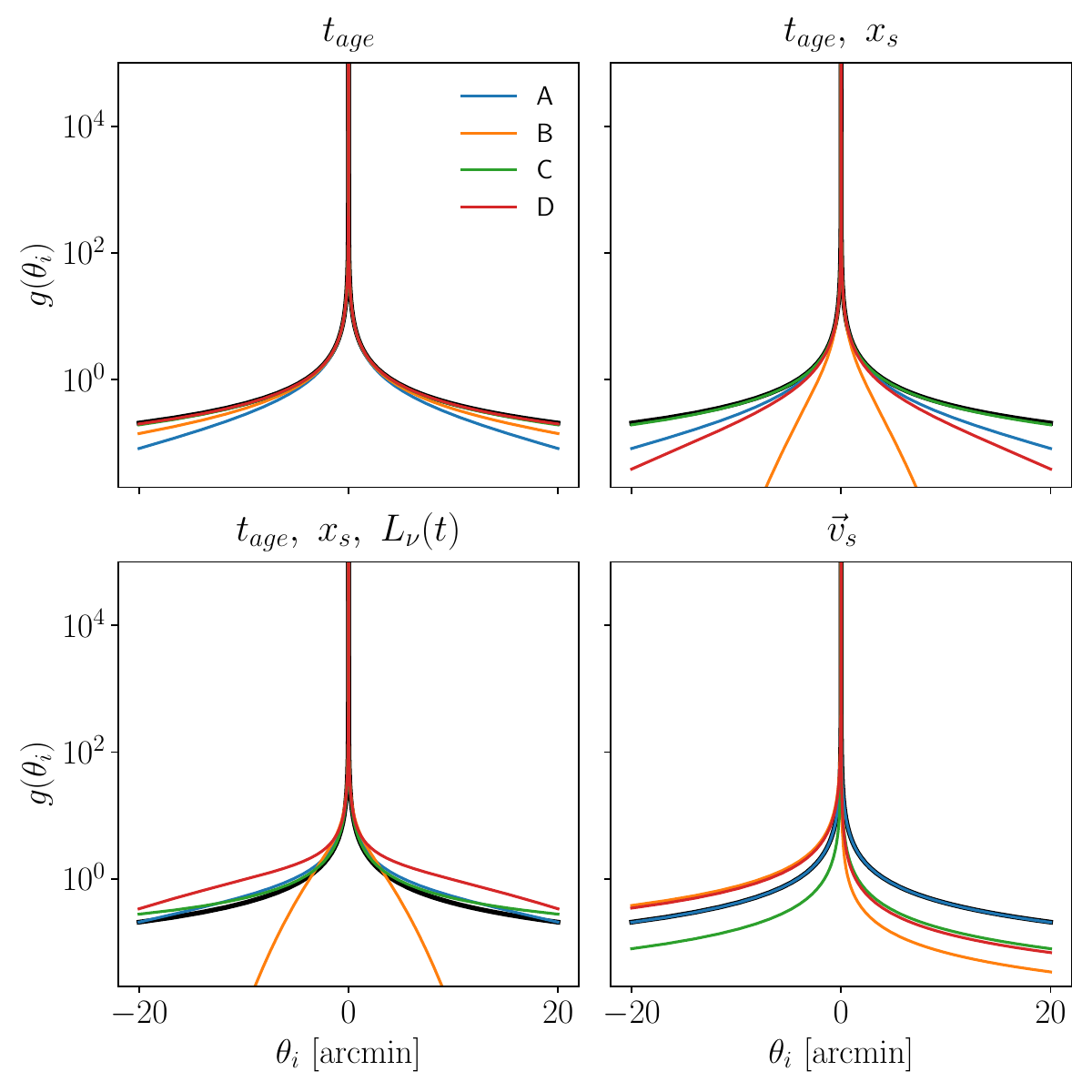}
\caption{Effects of the source characteristics on the front-light echo.}
\label{fig:sky_integrated_ideal_behind_source}
\end{figure}

In this Section, we study how the properties of the source affect the echo emission. In particular, we take into account the following ones:
\begin{enumerate}
    \item the age of the source $\tau_{age}$,
    \item $\tau_{age}$ in combination with the distance of the source $x_s$,
    \item $\tau_{age}$ in combination with the time dependence of the flux,
    \item the motion of the source with respect to Earth.
\end{enumerate}
Results are shown in Figures~\ref{fig:sky_integrated_ideal_opposite_source} and~\ref{fig:sky_integrated_ideal_behind_source}.
The top left panels show the effect of a finite a $\tau_{age}$. The age of the source translates into a reduced upper integration limit in Eq.~\eqref{g}. The time it takes a photon to travel the distance $x_d + x_{ds}$ cannot exceed $\tau_{age}$. The upper integration limit becomes then
\beq
R 
% = \frac{1}{2}\frac{\tau_{age}^2-x_s^2}{\tau_{age} \pm  x_s \cos\theta_i}
\sim
\frac{1}{2}(\tau_{age} \mp x_s) \enspace.
\eeq
On the top left panel of Figures~\ref{fig:sky_integrated_ideal_opposite_source} and~\ref{fig:sky_integrated_ideal_behind_source}, we see the expected reduction of the flux for the younger sources. For the older sources, the upper integration limit cuts out locations from where the flux is already suppressed by a large value of $x_{ds}$, and the effect is negligible.

On the top right panel, we add a varying distance for the sources. 
For the back-light echo, a larger distance implies a reduced upper integration limit, resulting in a reduced flux from sources B and D compared to the top left panel. For the front-light echo, a large $x_s$ means a larger $R$, but also a larger $\theta_d$ for fixed $\theta_i$. For this reason, we see a dimming of the flux compared to the top left panel even if the upper integration limit is larger.

For the time dependence of the sources' luminosity, we assume a power law
\beq
L_\nu(t) \propto \left(1 + \frac{t}{\tau_0}\right)^{-\alpha}  \enspace,\label{lumi}
\eeq
where $t$ is the time since the birth of the source. A positive $\alpha$ means that the source was brighter in the past. 
% If we observe an echo photon today at $t=\tau_{age}$, we have
% \beq
% \frac{L_\nu(t_{em})}{L_\nu(\tau_{age} - x_s)} = \left(\frac{1 + \frac{\tau_{age}-x_d-x_{ds}}{\tau_0}}{1 + \frac{\tau_{age} - x_s}{\tau_0}}\right)^{-n}  \enspace.\label{L_tem}
% \eeq
The bottom left panel shows how the flux increases for $\alpha=1$ and $\tau_0 = 10^3$~yr. As expected, the effect is larger for the older sources.

Finally, the bottom right panel shows the aberration effect due to the motion of the source. 
Assuming that the source moves with a constant velocity $\v_s$, its position at time $t$ is
\beq
\r_s(t) =(v_{s,x},~v_{s,x},~v_{s,z})\,(t + x_s )\enspace,
\eeq
where now $t=0$ is the time today. Let's focus on a given decay location determined by the observation angle $\theta_i$ and the distance along the los $x_d$. The motion of the source deforms and rotates the triangle that has Earth, the source, and the decay location as vertices. From $\v_s$, we can obtain expressions for the time-dependent Earth-source distance $\tilde x_s(t)$, source-decay location distance $\tilde x_{ds}(t)$ and the direction from which the source's photon reach the decay location $\theta_s(t)$, $\phi_s(t)$.
Explicit expressions are given in Appendix~\ref{sec:motion}. 
% From $\tilde x_s(t)$, $\tilde x_{ds}(t)$, $\tilde\theta_i(t)$, we then can calculate $\tilde\theta_d(t_{em})$, the angle between the incoming and echo photons, which determines to what extent the decay was suppressed by the DM phase space distribution.

Let us first comment on the effect on the front-light echo shown in Figure~\ref{fig:sky_integrated_ideal_behind_source}.
We have chosen the velocity of the sources such that source A is moving towards us, meaning that it was further away in the past. In this case, the only effect is a mild increase in $x_s / x_{ds}$ and $\theta_s$ compared to the case of a source at rest. We find that this effect is negligible.

Source B is moving in the positive $x$-direction of Figure~\ref{fig:geom2}, meaning it was further down the $x$-axis in the past. If we start with a positive observation angle, at any time in the past the triangle was more open, because the angle between the los to the decay location and the los to the source was larger than today.
This leads to a suppression of the flux. On the other hand, if we start with $\theta_i<0$, the triangle closes as we go back in time, becoming a line when the source was along the los. This causes an increase in the flux. 

To understand if this effect is important, we estimate at which distance along the los the decay must happen for us to be receiving today the echo radiation stimulated by photons emitted when the source was along the line of sight. The enhancement of the flux will be larger if the los crossing happened recently so that the decay location is close and well within the halo.

Setting $v_{s,y}=0$ we see from Figures~\ref{fig:geom1} and~\ref{fig:geom2}, that the source was along the los when $\theta_s(t)=\theta_i$. Starting from a positive $\theta_i$, this happened at a time $t_{em} = \mp x_s \theta_i / v_{s,x} $, (see Eq.~\eqref{thetas_phis_of_t})
This tells us that the zero crossing of $\theta_i$ has occurred in the past for the front-light echo if $\theta_i$ and $v_{s,x}$ have opposite signs, the other way around for the back-light echo. The distance of the decay stimulated by photons emitted at $t_{em}$ is
\beq
x_d \sim  \pm\frac{x_s}{2} \left(\frac{\theta_i}{v_{s,x}} - 1\right)\enspace,
\eeq
which may well be within the DM halo.

Going back to Figure~\ref{fig:sky_integrated_ideal_behind_source}, for source C, $\v_s$ is perpendicular to the plane $xz$-plane. This causes the triangle to open up, causing an overall reduction of the flux.
Finally, source D is moving both away from us and in the positive $x$-direction. The result is similar to that of source B, the effect being reduced because of a smaller magnitude of $v_{s,x}$.

Let's now turn to the back-light echo in Figure~\ref{fig:sky_integrated_ideal_opposite_source}. For sources B and D, we see that in this case, the flux is larger for positive $\theta_i$. This is expected because now, due to the different geometry, the source crossed the los in the past if $\theta_i>0$.

%%%%%%%%%%%%%%%%%%%%%%%%%%%%%%%%%%%%
%%%%%%%%%%%%%%%%%%%%%%%%%%%%%%%%%%%%
\section{Collinear emission}\label{sec:collem}

In this Section, we consider axion decays happening at locations between us and the source. In this case, the echo (with momentum $\k\,$) travels back toward the source, while the decay photons with the same momentum as the incoming ones (denoted by $\q\,$), travel toward us.

The Boltzmann equation now reads
\beqa
\frac{d^3\dot{n}_a}{dq^3}
&=& -
\frac{1}{(2\pi)^3}\frac{1}{2\omega_q}
\int \frac{d^3p}{(2\pi)^3}\frac{f_a(\p\,)}{2m_a}\ 
\frac{d^3k}{(2\pi)^3}\frac{1}{2\omega_{k}}
(2\pi)^4\delta^{(4)}(k + q - p) \sum_{\lambda}|\mathcal{M}_0|^2
(1+f_\gamma(\q, \lambda)) 
\enspace.\nl\label{boltzmann_collinear}
\eeqa

Following Appendix~\ref{sec:collin}, one obtains
\beqa
\frac{d\dot{n}_a}{d\omega_q d\Omega_q}
 &=& -
n_a \frac{g_{a\gamma}^2m_a^2}{64\pi^2}\,\frac{N_{pol}}{2}\, f_\gamma(\omega_q)\delta(\Omega_q)\
 h_{col}(\omega_q, \x_{ds})  \enspace,
\eeqa
with
\beq
h_{col}(\omega_q, \x_{ds}) = \frac{1}{\sqrt{2\pi}\delta v}\ e^{-\frac{(\varepsilon - \langle v_\parallel\rangle)^2}{2\delta v^2}}\enspace.
\eeq
The decay rate is the largest for $\varepsilon = \langle v_\parallel\rangle$, where, as before, $\langle v_\parallel\rangle$ is the average axion velocity along the direction of the momentum of the photon coming towards us, $\q$. 

Here $\Omega_q$ is the angle between the direction of the stimulating photon and the direction of the photon emitted in the ALP decay.
The delta-function $\delta(\Omega_q)$ indicates that the collinear emission comes exactly from the direction of the source, and is not smeared over the sky by the axion momentum dispersion as it happens instead for the echo case.

Proceeding like in Section~\ref{sec:echo_signal}, we obtain the flux density
\beq
S_\nu = \frac{\pi g^2}{4m}\ S_\nu^{(0)}
\int d\Omega_i \,B(\Omega_i) \int dx_d\, \delta(\Omega_q)\, \rho_a(\x_{ds})\,\left(\frac{x_s}{x_{ds}} \right)^2\, 
h_{col}(\omega_q, \x_{ds})\enspace.\label{eq:flux_density_collinear}
\eeq
With simple geometrical considerations, one can see that, in the limit of small angles, $\theta_q=\theta_i\,(x_d+x_{ds})/x_{ds}$, from which it follows $ \delta(\Omega_q)= \delta(\Omega_i)\, x_{ds}^2/x_s^2$.
We can thus recast Eq.~\eqref{eq:flux_density_collinear} in a simpler form:
\beq
S_\nu(t) = \frac{\pi g^2}{4m}\ S_\nu^{(0)}(t)
\,\int dx_{ds}\ \rho_a(\x_{ds})\,
h_{col}(\omega_q,  \x_{ds})\enspace.\label{flux_density_collinear_}
\eeq
where, in analogy to Section~\ref{sec:echo_signal}, we also generalized the equation to include the time dependence of the source's flux.

Compared to Eq.~\eqref{flux_density_gen}, the factor $L_\nu(t_{em})/L_\nu(t - \tilde x_s(t))$ is not present, because the photons produced in the decay travel together with the ones coming directly from the source, i.e. $t_{em} = t- \tilde x_s(t)$. Note that in the collinear case, if the source is time-dependent on short timescales, e.g., it has pulsations, the same time-dependency occurs also for the echo signal.

The motion of the source also enters, in principle, through a time-dependent $\delta(\Omega_i)$. However, as long as the source does not move out of the beam during the course of one observation, the proper motion of the source is completely negligible.

%%%%%%%%%%%%%%%%%%%%%%%%%%%%%%%%%%%%
%%%%%%%%%%%%%%%%%%%%%%%%%%%%%%%%%%%%
\section{Forecast for pulsars} \label{sec:forec}

Now, we apply the formalism we developed in the previous Sections to the case of Galactic pulsars as stimulating sources. 
Pulsars in connection to ALPs were considered, e.g.,~in \cite{Battye:2023oac}, but looking at a signal different from the one considered in our work. Ref.~\cite{Battye:2023oac} studied the ALP conversion in the pulsar magnetic field. 
Here we derive prospects for detecting the signals described in the previous Sections. We will focus on the front-light echo and collinear emission, which are peaked at the source location for pointlike sources like pulsars, unlike the back-light echo. Since the back-light echo is spread on angular sized of order 10' for both pointlike and non-pointlike stimulating sources, there is no advantage in considering pulsars rather than, for example, supernova remnants.

Our formalism applies to point sources emitting isotropically. The radio emission of pulsars is however not isotropic, but rather concentrated into a rotating beam confined within the polar cap region. The typical opening angle of the beam is~\cite{pulsarhandbook}
\beq
\rho
= 5.4^\circ \left(\frac{P}{\rm s} \right)^{-1/2}\enspace.
\eeq
Since the echo intensity is spread over angular distances of order tens of arcminutes, and barring complications due to possible changes of direction of the pulsar's rotation axis or magnetic field lines, we can safely consider the time-averaged pulsar emission to be isotropic in the directions that are relevant to us. 
A complication may arise due to the pulsar's proper motion. If the source was further away from the los in the past, it may be the case that the angle $\theta_s(t)$ was larger than $\rho$. In the following, we remove from our integration over the los locations for which this is the case. We find that this effect is completely negligible.

Since the flux density involves the product $S^{(0)}_\nu \,x_s\, g(\Omega_i)$, we focus on the pulsars with the largest flux times distance product in the Australia Telescope National Facility (ATNF) pulsar catalog~\cite{ATNF} and then take into account all the effects described in the previous Section.

Most of the relevant information is available directly from the ATNF catalog, with the exception of the source's velocity along the los and the time dependence of the source luminosity. 
From the analysis of Section~\ref{sec:source}, we know that the velocity along the los only affects the flux in a negligible way. We thus set it to zero.
The time dependence of the flux luminosity is discussed in the next sub-Section.

%%%%%%%%%%%%%%%%%%%%%%%%%%%%%%%%%%%%
%%%%%%%%%%%%%%%%%%%%%%%%%%%%%%%%%%%%
\subsection{Modeling of the time dependence of the pulsar flux}
The flux from a pulsar typically varies on timescales of order 1 second because of the pulsar's rotation. However, as already noted in~\cite{2022PhRvD.105f3007S}, this time dependence is not present in the echo flux. In fact, the echo is the aggregate of photons produced in decays that happened at different distances along the los. At these locations, the flux from the source peaks at different times, due to the different travel times from the source location. 
Since we integrate in Eq.~\eqref{flux_density_gen} over distances of order the size of the halo, the travel time is much longer than the pulsar period and thus we expect the echo flux to be constant in time. %over time scales of order the pulse period.
Thus, in the following, we consider the pulsar's luminosity averaged over timescales much larger than its period, but much smaller than its age.
On the other hand, the time dependence of collinear emission is exactly the same as that of the source's emission.

Pulsars are the brightest at their birth, their luminosity decreasing in time following the spin-down power, $\dot{E}$, evolution.
For the time-dependent evolution of the spin-down power in the magnetic dipole model (braking index $n=3$), we consider the following equation, e.g.~\cite{Orusa:2021tts}
\begin{equation}
\dot{E}(t)=\dot{E_0}\left(1 + \frac{t}{\tau_0}\right)^\beta  \enspace,\label{spindown_lum}
\end{equation}
with $\beta = -2$.
The constant $\tau_0$ is given by
\begin{equation}
    \tau_0 = \frac{P_0}{2 \dot{P_0}}\enspace,
\end{equation}
where $P_0$ and $\dot{P_0}$ are the initial spin and spin derivative of the pulsar.
It is very challenging to infer the initial $P_0$ and $\dot{P_0}$ distributions. For example, given $P$ and $\dot{P}$ today, it is not possible to determine $P_0$ unless an independent measure of the pulsar's age is available~\cite{pulsarhandbook}.

The authors of Ref.~\cite{Igoshev:2022uog} analyzed a population of young neutron stars associated with supernova remnants, for which it is possible to estimate the actual age of the pulsar. They found a log-normal distribution for the initial period, with mean $\log_{10}(P_0/ \rm s) = -1.04^{+0.15}_{-0.2}$ and standard deviation is $\sigma_{P_0} = 0.53^{+0.12}_{-0.08}$. 
In the dipole model, the magnetic field is constant, since it depends only on the product $P \dot{P}$, so that we can assume that the initial magnetic field distribution corresponds to the present one. From~\cite{Igoshev:2022uog}, the present B-field distribution is best-fitted by a log-normal with mean $\log_{10}(B/ \rm G) = 12.44$ and standard deviation $\sigma_B = 0.44$.
We use the mean values of these distributions as benchmarks to evaluate the impact of the pulsar spin-down temporal evolution.

Using the mean values of the initial spin and magnetic field distributions in expression for the characteristic magnetic field~\cite{pulsarhandbook}
\beq
B = 3.2\times 10^{19}~\mathrm{G} ~\sqrt{\frac{P}{\mathrm{s}}\dot{P}}\enspace,
\eeq
we obtain $\tau_0 = 1.79\times 10^4$~yr.

The next ingredient is the relation between the spin-down power and the pulsar's luminosity at a given frequency.
Recently, Ref.~\cite{Posselt:2022vrk}
presented the largest, uniform, census of young pulsars deriving for 1170 objects correlations between the radio (pseudo)luminosity and parameters such as age and spin-down power. 
They found that, in the L-band, $\log L_{L} = c_{\dot{E}} \log\dot{E}$ with $c_{\dot{E}} = 0.15 \pm 0.02$.
Assuming $c_{\dot{E}}$ to be independent, we obtain that the luminosity of a pulsar follows Eq.~\eqref{lumi} with exponent $\alpha=0.3$.

We find that with this value of $\tau_0$ and $c_{\dot{E}}$ the effect on the echo of time dependence of the source's flux is negligible. We thus neglect this effect in our forecasts. In doing this we are being conservative, since, for example, the actual value of $\tau_0$ for a given pulsar may very well be smaller than our benchmark value.

%%%%%%%%%%%%%%%%%%%%%%%%%%%%%%%%%%%%
%%%%%%%%%%%%%%%%%%%%%%%%%%%%%%%%%%%%
\subsection{Target selection} \label{sec:target}
\begin{figure}
  \centering
  \includegraphics[width=\textwidth]{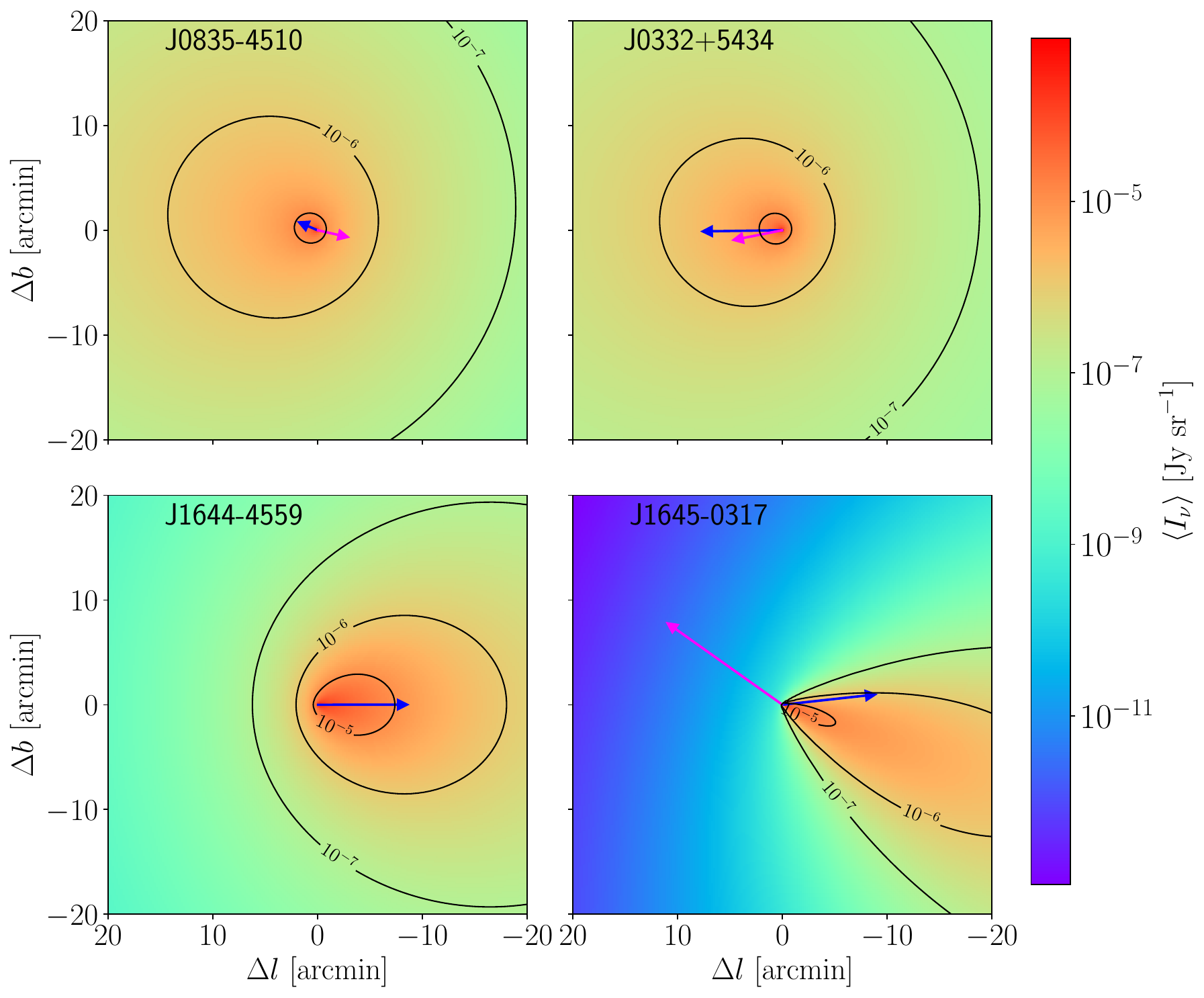}
\caption{Bandwidth-averaged front-light echo spectral intensity at 400~MHz for four of the most promising pulsars from the ATNF catalog assuming $g_{a\gamma}=10^{-11}~\mathrm{GeV}^{-1}$. Blue arrows indicate the direction of the component of the average DM velocity perpendicular to the line of sight, while magenta arrows indicate the source's proper motion. The arrows are drawn in arbitrary units, but their relative lengths are to scale with each other. The thin black lines mark contours of constant $\langle I_\nu\rangle$.}
\label{fig:sky_intensity_2D}
\end{figure}

\begin{figure}
  \centering
  \includegraphics[width=\textwidth]{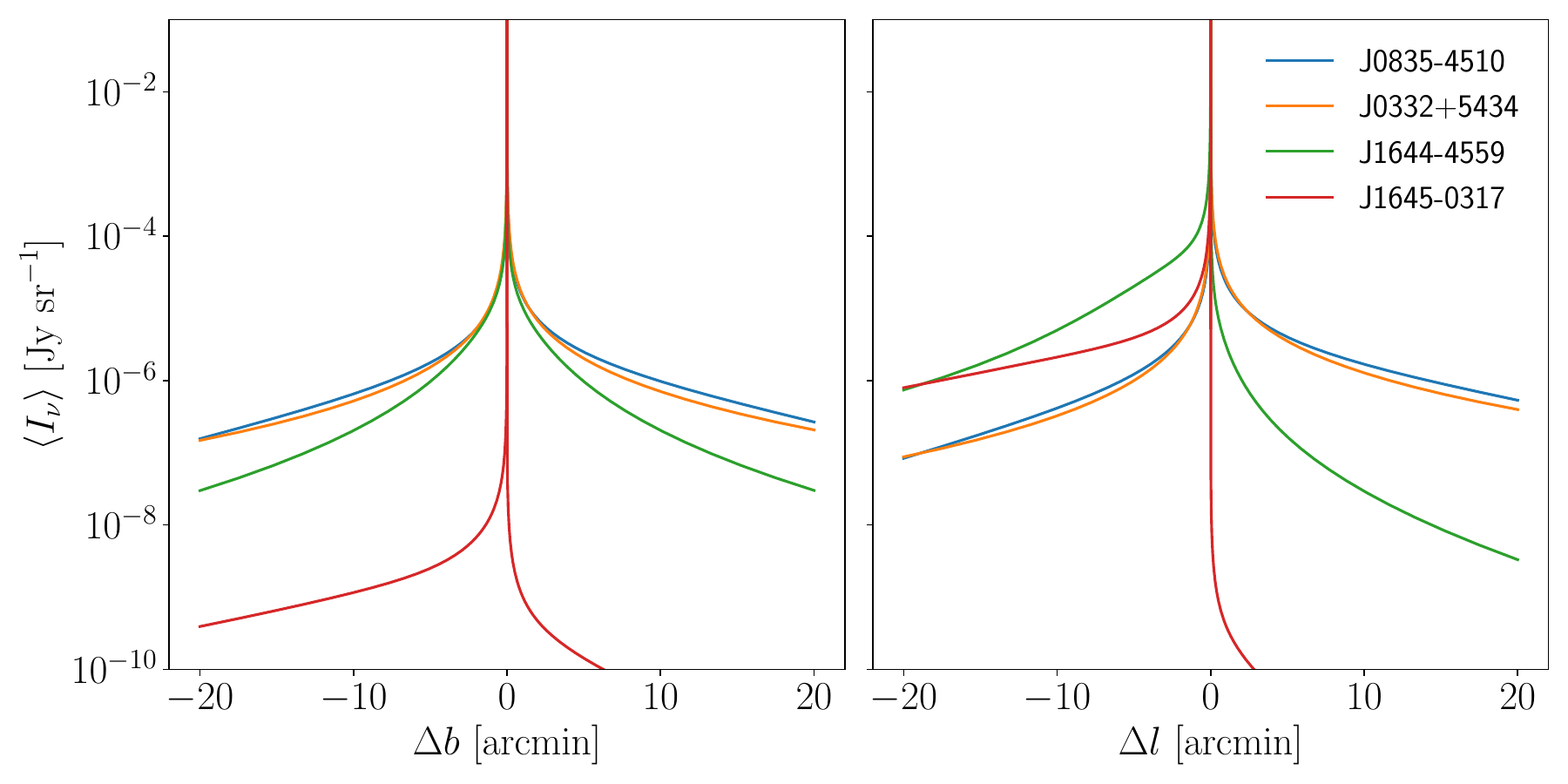}
\caption{Bandwidth-averaged front-light echo spectral intensity at 400~MHz for four of the same pulsars as Figure~\ref{fig:sky_intensity_2D} assuming $g_{a\gamma}=10^{-11}~\mathrm{GeV}^{-1}$. On the left (right) panel, we vary the Galactic latitude (longitude) of the line of sight, keeping the Galactic longitude (latitude) fixed.}
\label{fig:sky_intensity}
\end{figure}

To estimate our projected sensitivity, we choose to integrate over the bandwidth that maximizes the signal-to-noise ratio. Assuming noise scales like $\sqrt{\Delta\nu}$, a top-hat bandwidth and $\nu=m_a (1 + \langle v_\parallel\rangle)/4\pi$, we find that the optimal bandwidth is $\Delta\nu/\nu = 2.8 \delta v$~\cite{2022PhRvD.105g5006B}. 

For the targets we are considering, the effect on the echo of a position-dependent velocity dispersion is negligible. We fix then $\Delta\nu$ based on the velocity dispersion at Earth.
The frequency-averaged echo spectral intensity is
\beq
\langle I_\nu\rangle = 
f_{\Delta\nu}\,
\frac{g_{a\gamma}^2}{16\Delta\nu}\ S_\nu^{(0)} \int dx_d\ \rho_a(\x_{ds})\, \left(\frac{x_s}{x_{ds}} \right)^2
h_1(\omega_k, \x_{ds})\enspace,\label{spectral_intensity}
\eeq
with
\beqa
h_1(\x_{ds})
= \frac{1}{8\pi\delta v^2}
\ e^{-\frac{\langle v_t\rangle^2}{2\delta v^2}}
\ e^{-\frac{(\theta_d/2 - \langle v_\perp\rangle)^2}{2\delta v^2}}
 \enspace,
\eeqa
where we have neglected terms of order $\theta_d^2$ or $v\theta_d$. The Gaussian integrated over the chosen bandwidth provides $f_{\Delta\nu} = 0.84$. The frequency-averaged flux density can be then obtained from the usual $\langle S_\nu \rangle = \int d\Omega_i \,B(\Omega_i)\int_{\rm los} d\langle I_\nu \rangle $.

Similarly, for the collinear emission we have:
\beq
\langle S_\nu\rangle = 
f_{\Delta\nu}\frac{g_{a\gamma}^2}{16\Delta\nu}\ S_\nu^{(0)}
\int dx_d\ \rho_a(x_d)\enspace.\label{spectral_intensity_collinear}
\eeq

In Figure~\ref{fig:sky_intensity_2D}, we show the bandwidth-averaged front-light echo spectral intensity at a frequency of 400~MHz for the four pulsars in the ATNF catalog with the largest product of distance times flux. We assume an axion photon coupling $g_{a\gamma}=10^{-11}~\mathrm{GeV}^{-1}$. We take the flux from the best fit to observations from the recent catalog of~\cite{Swainston:2022gkj}. We note that the flux at 400~MHz is slightly lower than the one reported in ATNF. 
In Figure~\ref{fig:sky_intensity}, we show slices of constant Galactic longitude and latitude to ease the comparison among pulsars.

All effects discussed in Section~\ref{sec:realistic_case} are implemented with the exception of the time dependence of the source's luminosity.
%, including the non-trivial combination between $t_{age}$ and $\v_s$, and $L(t)$ and $\v_s$.
We have set $v_{s,z}=0$, as this quantity is not known and we have established it is not important for the echo.
The age of the pulsar used to determine the upper limit of integration is taken to be the characteristic age $\tau_{age}$ plus the pulsar distance.

The Vela pulsar J0835-4510 is the brightest source at this frequency with a flux of about 5~Jy. Its front-light echo flux does not experience significant reduction due to any of the effects discussed in Section~\ref{sec:realistic_case}.
There is some dimming compared to the idealized case due to the young age $\tau_{age} = 1.13\times 10^4$~yr, and, to a lesser extent, to the DM density along the los. The DM velocity is essentially along the los, thus affecting the resonance frequency, but not the distribution of the echo on the sky. The projection on the sky of $\langle \v\rangle$ is shown by the blue arrow in Figure~\ref{fig:sky_intensity_2D}.
The small velocity of the source, $\v_s \sim 60$~km/s, shown by the magenta arrow,  makes the intensity larger for positive values of $\Delta l$, as also visible on the right panel of Figure~\ref{fig:sky_intensity}.

Pulsar J0332+5434 has the largest $x_s\,S_\nu^{(0)}$, a factor 1.8 larger than Vela.  However, its Galactic longitude $l=144^\circ$, and distance $x_s = 1.695$~kpc, imply that the average DM density along the los is low making it lose its ``source advantage". 
The average DM and source velocities point approximately in the same direction. From the discussion of Section~\ref{sec:halo}, we know that for the front-light echo, the intensity is larger in the direction of $\langle \v\rangle$, while in Section~\ref{sec:source}
we have learned that the intensity is larger in the direction of $-\v_s$ (where the source was in the past). For J0332+5434 the two effects tend to compensate, resulting in an intensity patch similar to that of Vela.

Next, we have J1644-4559 and J1645-0317. These two pulsars have similar fluxes at 400~MHz of about 0.4~Jy and comparable distances of order 4~kpc.
Their Galactic coordinates are $(339^\circ,~ -0.195^\circ)$ and $(14.1^\circ,~26.1^\circ)$, respectively, implying that the DM density along both lines of sight is comparable, as well as the projection of the DM velocity on the sky. The main difference between the two is due to our description of their proper motion. For J1644-4559, proper motion information is not available and we set its velocity to zero. J1645-0317 on the other hand has a large velocity of about 375~km/s which causes a drastic reduction in the flux in the direction of $\v_s$.

This discussion highlights the importance of our careful treatment of the signal, which we demonstrate to vary on a source-by-source basis.
Different targets therefore reveal a very diverse signal phenomenology.

%%%%%%%%%%%%%%%%%%%%%%%%%%%%%%%%%%%%
%%%%%%%%%%%%%%%%%%%%%%%%%%%%%%%%%%%%
\subsection{Observational prospects}\label{sec:prospects}
\begin{figure}
  \centering
  \includegraphics[width=0.8\textwidth]{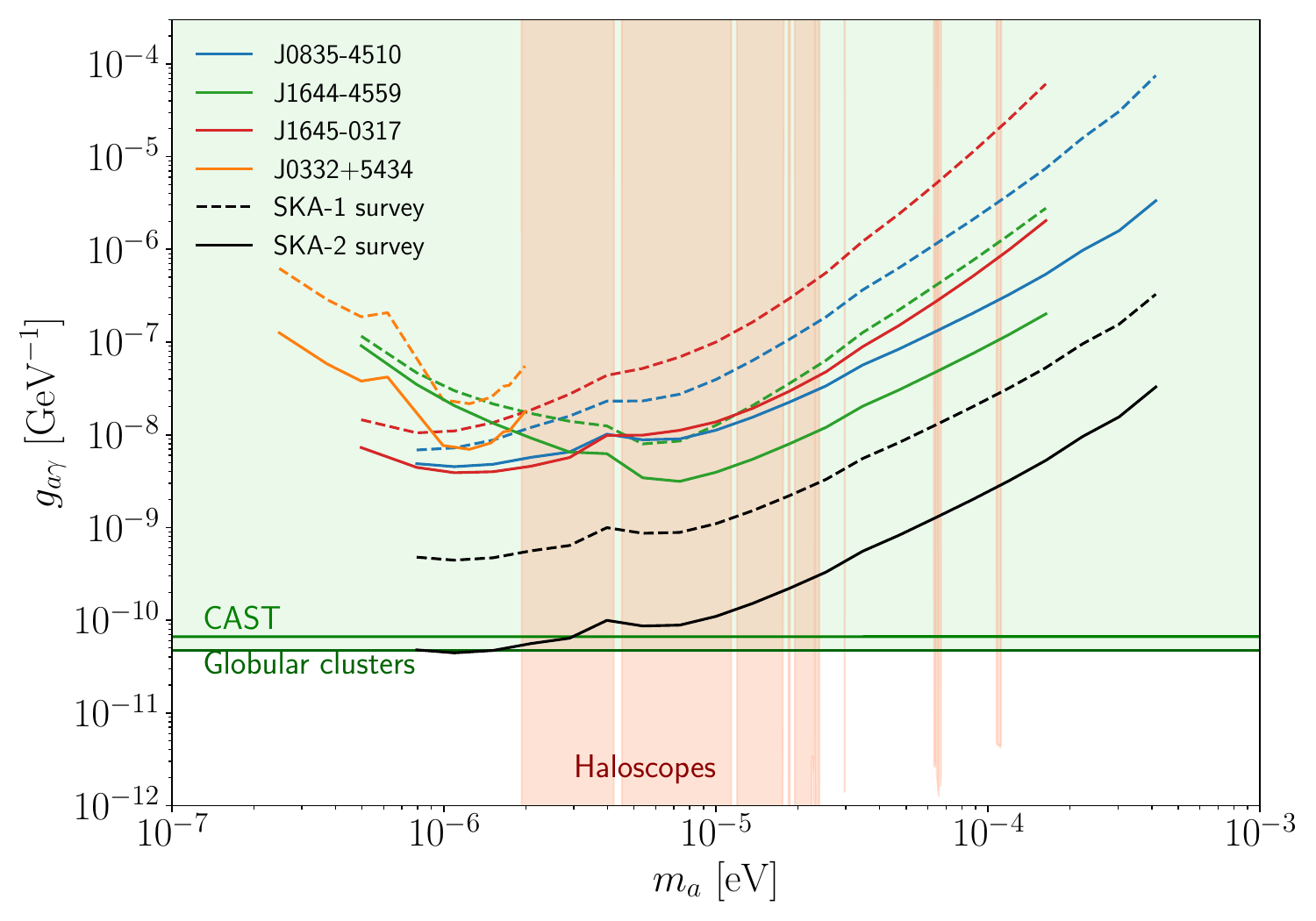}
\caption{Projected sensitivity on the ALP-photon coupling $g_{a\gamma}$ versus ALP mass $m_a$, for different scenarios, described in Section~\ref{sec:prospects}. The upper colored cases correspond to observations of single pulsars with either SKA-1 or LOFAR 2.0. More precisely, we show the cases of J0835-4510 (blue), J1644-4559 (green), J1645-0317 (red), and J0332+5434 (orange), including both the front-light echo signal (dashed) and collinear emission (solid). The two black curves report the sensitivity for surveys of all Galactic pulsars with SKA-1 (dashed) and SKA-2 (solid) considering only the collinear case.}
\label{fig:forecast}
\end{figure}

For our forecasts, we consider two interferometric observatories of the near future: SKAO for the Southern hemisphere and LOFAR 2.0 for the Northern hemisphere.

Concerning the performance of the first phase of the SKA project, which used to be referred to as SKA-1, we follow \cite{SKA:2019}, see their Tables 6 and 7. For example, at 114 MHz, the sensitivity of SKA-1 Low is $\sigma\sim 184\,\mu{\rm Jy}/{\rm beam}\,\sqrt{100\,{\rm hr}/\Delta t}\,\,\sqrt{10^{-4}/(\Delta \nu/\nu)}$ for a beam of size between $12^{\prime\prime}$ and $10.2^\prime$, while for SKA-1 Mid at 1.05 GHz, the sensitivity is $\sigma\sim 18.6\,\mu{\rm Jy}/{\rm beam}\,\sqrt{100\,{\rm hr}/\Delta t}\,\,\sqrt{10^{-4}/(\Delta \nu/\nu)}$ for a beam of size between $0.8^{\prime\prime}$ and $1.8^\prime$.
SKA-1 should become operative in its final configuration in 2029.

A further development of the SKAO, which used to be referred to as SKA-2, is foreseen for the 2030's, with telescope specifications still to be decided. According to~\cite{SKA2}, the goal is to reach a factor of $\times 10$ improvement in sensitivity with respect to SKA-1. 

For very low-frequency emission and/or Northern hemisphere targets, we consider LOFAR 2.0, which is an ongoing program including a set of upgrades of the LOFAR stations and an increase of the capabilities of the array, expected to be completed in 2024.
We consider the foreseen sensitivity reported in \cite{LOFAR2}.
At 150 MHz, it is $\sigma\sim 200\,\mu{\rm Jy}/{\rm beam}\,\sqrt{100\,{\rm hr}/\Delta t}\,\,\sqrt{10^{-3}/(\Delta \nu/\nu)}$ for a beam at arcsec resolution.

In this Section, we will derive forecasts for the front-light echo and collinear emission. 
For both of them, we can use the point-like source sensitivity of the telescopes. Bounds are derived requiring $\langle S_\nu \rangle<2 \sigma_{rms}$, and assuming that the background continuum emission in a single frequency channel can be subtracted away by fitting its shape in the full bandwidth.
In the collinear case, we consider an improvement in the sensitivity by a factor $\sqrt{W / (P-W)}$, where $W$ is the width of the pulsation and $P$ is the period.

The first case shown in Figure~\ref{fig:forecast} concerns the sensitivity forecast for 100~hr observations of single pulsars (colored curves).
We consider the same four pulsars described in Section~\ref{sec:target}.
The pulsars J0835-4510, J1644-4559, and J1645-0317 are in the Southern hemisphere, under the SKA-1 reach. For the observation of J0332+5434 in the Northern hemisphere we instead use LOFAR 2.0.
In the case of the front-light echo (color dashed), we remove the contribution from observation angles smaller than the frequency-dependent size of the broadened pulsar image. This would be given by photons passing again through the pulsar, where they might be absorbed or rescattered. For simplicity, we discard those photons. The pulsar's apparent angular size is given by the broadening due to the scattering of photons in the interstellar medium and it is related to the scattering timescale $\tau_s$ as $\theta_b = \sqrt{\tau_s/x_s}$~\cite{pulsarhandbook}.

As it can be seen from Figure~\ref{fig:forecast}, in this case, prospects fall into a strongly constrained region of the ALP parameter space, already excluded by the CAST helioscope results. We find that the collinear emission provides the best sensitivity.

On top of the forecast for a single pulsar, we then estimate the projected sensitivity in the case of a survey of all Galactic pulsars. Such survey is planned for SKAO~\cite{Smits:2009} with an already ongoing program along this direction conducted with MeerKAT~\cite{Song:2021}.
Providing an accurate sensitivity forecast for the survey case is beyond the goal of this paper since it would require knowing the schedule of observations and performing simulations of pulsar populations. We can however derive a simple but reliable projected sensitivity in the following way.
Let's assume Vela (J0835-4510) to be a typical pulsar, in terms of luminosity and other properties, and to be the closest one, which also means the brightest one in terms of flux. We consider a radio luminosity distribution $N\propto L^{-0.8}$ for Galactic pulsars, as found in \cite{Lorimer:2006}. Let us now take a conservative simple 1D distribution along the direction to the Galactic center, or, say in other words, let us conservatively assume $N\propto S^{(0)-0.8}_\nu$ (in reality it will be somewhat steeper). This implies that the collective pulsar flux measured in a survey is $\int dS\,S\,dN/dS\simeq 4\,S^{(0)}_{\rm Vela} $. Then we take an average pulsar distance of $\langle d\rangle \simeq 8$ kpc and an average DM density between us and the pulsar of $\simeq 2.9\,\rho_\odot $, obtained volume averaging the Navarro-Frenk-White profile used in Section~\ref{sec:sky} above in the inner 8 kpc from the Galactic center. Finally, we assume 10 hr of observations for each survey field (which, in general, will contain several pulsars), instead of 100 hr considered for the single pulsar case discussed above. 
Combining all these arguments we obtain an improvement of a factor of $\sim 10.2$ in the projected $g_{a\gamma}$ bound with respect to the case considering the Vela pulsar only.

In Figure~\ref{fig:forecast}, we show the forecast for a survey with SKA-1 assuming this simple rescaling factor (black dashed). We focus on the collinear signal since we saw from the single pulsar case that it is more promising than the front-light echo.
%For LOFAR 2.0, one can follow the same argument, now starting from the results obtained for J0332+5434 instead of Vela.

Our ultimate forecast is for a survey with SKA-2 (black solid). With such a strategy, we will be able to touch the more interesting region of the ALPs parameter space, providing complementary bounds to haloscopes. A detailed forecast making use of pulsars' population models and more precise survey details is beyond the scope of the present work.

\section{Conclusions}\label{sec:conc}
In this work, we present the in-depth formalism for computing the signals associated with photon emission from the stimulated decay of axions in the Galactic halo. We started from first principles and focused on stimulating point-like sources in our Galaxy. The presence of $\mu$eV ALPs in the Milky Way halo would imply the existence of three distinct line signatures:
\begin{enumerate}
    \item {\bf Front-light echo}: Stationary emission from approximately the same direction of the stimulating source and with a slightly broadened size and an orthogonal polarization.
    \item {\bf Back-light echo}: Stationary emission from the opposite direction with respect to the stimulating source and with a significantly broadened size and an orthogonal polarization.
    \item {\bf Collinear emission}: Possibly pulsed emission (if the stimulating source is pulsed) with the same direction, size, and polarization of the stimulating source.
\end{enumerate}
While only the back-light echo has been discussed in the literature so far, we here 
 highlight the co-existence of these three line signatures with the above properties.\footnote{Once this work was in its final phase, Ref.~\cite{2023arXiv231003788S} appeared on arXiv. They discuss the two echoes (called gegenschein) and the collinear emission (called forwardschein), focusing on the observational prospects associated with supernova remnants. Our work aims instead at providing a detailed and general derivation of the effects.}
It is difficult to imagine an astrophysical source providing simultaneously all of them.
Therefore, they offer a unique opportunity for a clear ALP identification.

Previous works~\cite{Caputo:2018ljp,Caputo:2018vmy,Arza:2019nta,2022PhRvD.105b3023A,2023arXiv230906857A,2023PhRvD.108h3001A,2020arXiv200802729G,2022PhRvD.105g5006B,2022PhRvD.105f3007S} have discussed similar or connected signals, and here we aim at providing a comprehensive view, and an in-depth discussion of the rich and diverse phenomenology that arises from different targets.
In particular, we discussed how the echoes and the collinear emission depend on the ALP DM
properties, namely on the DM spatial and velocity distributions, and on the source characteristics, such as distance, age, and motion. These dependencies offer extra handles for the ALP signal identification.

In the final part of the paper, we focused on the case of Galactic pulsars as stimulating sources.
We showed that pulsar radio surveys of the next decade can be used to identify the described signatures for $\mu$eV ALPs.

\section*{Acknowledgements}
We would like to thank M. Taoso for useful discussions.

MR and ET acknowledge support from the project ``Theoretical Astroparticle Physics (TAsP)'' funded by the INFN, from `Departments of Excellence 2018-2022' grant awarded by the Italian Ministry of Education, University and Research (\textsc{miur}) L.\ 232/2016, from the research grant `From Darklight to DM: understanding the galaxy/matter connection to measure the Universe' No.\ 20179P3PKJ funded by \textsc{miur}, and from the ``Grant for Internationalization" of the University of Torino.

ET thanks the Galileo Galilei Institute for Theoretical Physics for its hospitality during the initial stages of this work.

This article/publication is based upon work from COST Action COSMIC WISPers CA21106, supported by COST (European Cooperation in Science and Technology).

\bibliographystyle{JHEP_improved}
\bibliography{biblio}

%%%%%%%%%%%%%%%%%%%%%%%%%%%%%%%
%%%%%%%%%%%%%%%%%%%%%%%%%%%%%%%
%%%%%%%%%%%%%%%%%%%%%%%%%%%%%%%
\appendix

%%%%%%%%%%%%%%%%%%%%%%%%%%%%%%%%%%%%%%%%%%%%%%%%%%%%%%%%
%%%%%%%%%%%%%%%%%%%%%%%%%%%%%%%%%%%%%%%%%%%%%%%%%%%%%%%%
%%%%%%%%%%%%%%%%%%%%%%%%%%%%%%%%%%%%%%%%%%%%%%%%%%%%%%%%
\section{Derivation of the decay rate}\label{sec:signal}

In this Section, we derive the number of ALP decays per unit time and unit volume that produce photons of a given frequency $\omega_k$ traveling in a given direction $\Omega_k$, given the position of a point-like stimulating source that emits radially and isotropically. The number of decays can be related to the echo flux density at Earth as described in Section~\ref{sec:echo_signal}.

Our starting point is the Boltzmann equation Eq.~\eqref{boltzmann}. 
Throughout the following derivation, we assume ALPs are non-relativistic. We keep corrections in the ALP momentum $\p$ only up to linear order, and only when they are relevant to the echo's frequency or direction. Namely, when the correction only affects the magnitude of the decay rate, we ignore it.

Using the delta-function to eliminate the integral over $\q$ in Eq.~\eqref{boltzmann}, and keeping only the stimulated decay term, we obtain
\beqa
\frac{d^3\dot{n}_a}{dk^3}%
 &=& -
\frac{1}{(2\pi)^3}\frac{1}{(2\omega_k)^2}
 \int \frac{d^3p}{(2\pi)^3}\frac{f_a(\p\,)}{2m_a}\ 
 (2\pi)\delta(\omega_q + \omega_k - m_a) \sum_\lambda|\mathcal{M}_0|^2
 f_\gamma(\q, \lambda)\Big|_{\q=-\k+\p}
 \enspace.\nl\label{start_equation}
\eeqa
The matrix element is
\beq
\sum_\lambda|\mathcal{M}_0|^2 = g_{a\gamma}^2 \, |\epsilon^{\mu\nu\rho\sigma} (\hat{e}^{(1)}_\mu \hat{e}^{(2)}_\nu)^* k_{1,\rho} k_{2,\sigma}|^2 =  g_{a\gamma}^2 \frac{m_a^4}{2} \frac{N_{pol}}{2}\enspace,
\eeq
where we allow the possibility for the source to be polarized. Let us note here that the matrix element for axion decay into two photons is often reported with a symmetry factor $1/2$ for identical particles in the final state, which avoids double counting of identical photons in the $4\pi$ angular integration. This is for the isotropic case, whilst with our definition of the photon phase-space in Eq.~\ref{fgamma}, there is only one possible configuration and no double-counting.
The frequency of the incoming photon can be expanded as
\beq
\omega_q \sim \omega_k - p \cos\psi \enspace,
\eeq
where $\psi$ is the angle between $\k$ and $\p$.
% In polar coordinates, we have 
% \beq
% {\cos\psi = \cos\theta_{\k}\cos\theta_{\p} + \sin\theta_{\k}\sin\theta_{\p} \cos(\phi_{\k}-\phi_{\p})}\enspace.
% \eeq
From the energy-conserving delta function, we see that the frequency of the echo photons must be close to half the ALP mass, shifted by a small amount proportional to the ALP velocity along the direction of $\k$
\beq
\omega_k = \frac{m_a + p\cos\psi}{2}\enspace.
\eeq

We choose the coordinate system of Figures~\ref{fig:geom1} and~\ref{fig:geom2} and focus on a point source emitting radially and isotropically by using Eq.~\eqref{fgamma}.
Keeping in mind that $f_\gamma$, $f_a$, and thus $d^3\dot{n}_a /dk^3$ depend on the decay location $\x_{ds}$, we can write
\beqa
\frac{d^3\dot{n}_a}{dk^3}
&=& - 
 \frac{g_{a\gamma}^2 m_a}{4} \frac{N_{pol}}{2}\, f_\gamma\left(\frac{m_a}{2}\right)
\frac{1}{(2\pi)^5}\ I(\k\,) \enspace,
\eeqa
with
\beqa
I(\k\,) &=&\int d^3p\  f_a(\p\,)\ 
\delta(\omega_q + \omega_k - m_a) 
 \frac{\delta(\theta_{q} - \theta_s)\delta(\phi_{\q})}{\sin\theta_{\q}}\Big|_{\q=-\k+\p}
 \enspace,\label{help_integral}
\eeqa

To eliminate the integrals over the axion phase space, we need to relate the angles of $\q$ to those of $\p$. 
We choose the decay location to be positioned in the first quadrant on the $xz$-plane. Then the momentum of the echo photon must have an azimuthal angle  $\phi_{\k} = \pi$.
Requiring that $\phi_{\q} = 0$ then yields $\phi_{\p} = 0, \pi$.
Solving $\p = \k +\q$, we get 
\beqa
p &=& \sqrt{\omega_k^2 + \omega_q^2 + 2 \omega_k\omega_q\cos(\theta_{\k}+\theta_{\q})}\nl
\theta_p &=&
\begin{cases}
-\arctan\left(
\frac{
\omega_k\cos\theta_{\k} + \omega_q\cos\theta_{\q}
}{
\omega_k\sin\theta_ k- \omega_q\sin\theta_{\q}
}\right), & \text{if $\phi_p=0$}\\
\arctan\left(
\frac{
\omega_k\cos\theta_{\k} + \omega_q\cos\theta_{\q}
}{
\omega_k\sin\theta_ k- \omega_q\sin\theta_{\q}
}\right), & \text{if $\phi_p=\pi$}\\
\end{cases}
\eeqa
We can now perform the change of variable $d^3p\to d^3q$ in Eq.~\eqref{help_integral}
\beqa
I (\k\,)
&=& 
\omega_k^2\int d\omega_q\, d\theta_q\,d\phi_q \ f_a(\p\,)\ 
\delta(\omega_k+\omega_q - m) 
\delta(\theta_{\q} - \theta_s)\delta(\phi_{\q})\Big|_{\p=\p(\q\,)} \nl
&=& 
\omega_k^2\ f_a(\p\,(\q\,))\ 
\Big|_{\substack{\omega_q = m-\omega_k \\ \theta_{\q} = \theta_s}}\enspace.
\eeqa
The equation above tells us that, for a fixed direction of stimulating photons arriving at the decay location, given by $\theta_s$, an ALP with a specific momentum needs to be involved in the decay in order for the echo to have momentum $\k$ and to reach Earth.
In the coordinate system of Figures~\ref{fig:geom1} and \ref{fig:geom2}, the components of the axion momentum involved in such a decay are
\beqa
p_x &=& -\omega_k \sin\theta_{\k} + (m - \omega_k) \sin\theta_s \nl
p_y &=& 0\nl
p_z &=& \omega_k \cos\theta_k + (m - \omega_k) \cos\theta_s \enspace.\label{selected_p}
\eeqa
It is convenient to express $\theta_s$ in terms of the angle between the direction of the photons from the source coming into the decay location and the direction of the echo photons traveling towards Earth
\beq
\theta_d = \pi - \theta_{\k} - \theta_s\enspace. \label{thetad_def}
\eeq
From Figures~\ref{fig:geom1} and ~\ref{fig:geom2}, we see that $\theta_d >0$ for the back-light echo, while $\theta_d < 0$ for the front-light echo. 
One can check that $\theta_d$ is of order the axion velocity. More precisely, to linear order in $p$
\beqa
\theta_d &=& \frac{p}{\omega_k}\sin(\theta_{\k} \pm \theta_{\p})\label{thetakp}\enspace. 
\eeqa

We can then expand Eqs~\eqref{selected_p} to first order in the small quantities $\theta_d$ and $\epsilon = 2\omega_k -m$, obtaining
\beqa
p_x &=& \omega_k \theta_d   \ \cos\theta_k - \epsilon\sin\theta_k\nl
p_z &=& \epsilon\ \cos\theta_k + \omega_k \theta_d \ \sin\theta_k\enspace.\label{pxpz}
\eeqa

Let's define components of the axion momentum parallel and perpendicular to $\k$, through the following rotation
\beq
\left(
\begin{array}{cc}
p_\perp\\
p_\parallel \\
\end{array}
\right)
=
\left(
\begin{array}{cc}
 \cos\theta_k & \sin\theta_k \\
 -\sin\theta_k & \cos\theta_k \\
\end{array}
\right)
\left(
\begin{array}{cc}
p_x\\
p_z \\
\end{array}
\right)
=
\left(
\begin{array}{cc}
\omega_k \theta_d\\
\epsilon \\
\end{array}
\right)\enspace.
\label{rotation_xz}
\eeq
The axion momentum $\p\,$ necessary to produce an echo photon with frequency $\omega_k$ and direction $\theta_k$, given the direction of the incoming photon $\theta_s$,
has a component along the direction of $\k$ equal to $\epsilon$, and the component perpendicular to $\k$ equal to $\omega_k \theta_d$.

Inserting our expression for $\p$ into Eq.~\eqref{fa}, we obtain the desired result
\beqa
I(\k\,)
&=& n_a \,
\left(\frac{m_a}{2}\right)^2
\frac{(2\pi)^3}{(2\pi)^{3/2}\delta p^3}
\ e^{-\frac{\langle p_t\rangle^2}{2\delta p^2}}
\ e^{-\frac{(\epsilon - \langle p_\parallel\,\rangle)^2}{2\delta p^2}}
\ e^{-\frac{(\omega_k\theta_d - \langle p_\perp\,\rangle)^2}{2\delta p^2}}
 \enspace,\label{d3ndk3_app}
\eeqa
where $ \langle p_\parallel\,\rangle$ and $ \langle p_\perp\,\rangle$ are the components of the axion average momentum parallel and perpendicular to $\k$ in the $xz$-plane, while $\langle p_t\rangle$ is the component perpendicular to the $xz$-plane.
Finally, we need to express the angle $\theta_d$ in terms of $x_d$, $x_s$ and $\theta_i$.
In the case of the back-light echo, we have
\beq
\cos\theta_d = \frac{x_d + x_s \cos\theta_i}{x_{ds}}\qquad\qquad
\sin\theta_d = \frac{x_s}{x_{ds}} \sin\theta_i \qquad\qquad
x_{ds} =  \sqrt{x_d^2+x_s^2 + 2x_d x_s\cos\theta_i}
\enspace.
\eeq
In this case, a small $\theta_i$ always implies a small $\theta_d$ and we can use
\beq
\theta_d = \frac{x_s}{x_{ds}} \theta_i \label{thetad_small}\enspace.
\eeq
For the front-light echo, we have instead
\beq
\cos|\theta_d| = \frac{x_d - x_s \cos\theta_i}{x_{ds}}\qquad\qquad
\sin|\theta_d| = \frac{x_s}{x_{ds}} \sin\theta_i\qquad\qquad
x_{ds} =  \sqrt{x_d^2+x_s^2 - 2x_d x_s\cos\theta_i} \enspace.
\eeq
In this case, $|\theta_d|$ can be large even if $\theta_i$ is small. In particular, $|\theta_d| > \pi/2$ for $\cos\theta_i > x_d/x_s$. However, such large values of $\theta_d$ are exponentially suppressed by the axion phase space distribution and won't contribute to the signal. For $\cos\theta_i \leq x_d/x_s$, we can express $\theta_d$ as
\beq
\theta_d = -\arcsin\frac{x_s}{x_{ds}} \theta_i \enspace.
\eeq

%%%%%%%%%%%%%%%%%%%%%%%%%%%%%%%%%%%%%%%%%%%%%%%%%%%%%%%%
%%%%%%%%%%%%%%%%%%%%%%%%%%%%%%%%%%%%%%%%%%%%%%%%%%%%%%%%
%%%%%%%%%%%%%%%%%%%%%%%%%%%%%%%%%%%%%%%%%%%%%%%%%%%%%%%%
\section{Motion of the source}\label{sec:motion}

Let's now take into account the fact that, in general, the source moves relative to us with some velocity $\v_s$. 
This implies that, for a fixed decay location $\x_{ds}$, the direction of the incoming photons from the source is time-dependent. The photon phase space density Eq.~\eqref{fgamma}
generalizes to
\beqa
f_\gamma(\k, \x_{ds}) &=& f_\gamma(\omega, x_{ds})
\frac{\delta(\theta_{\k} - \theta_s(t))\delta(\phi_{\k} - \phi_s(t))}{\sin\theta_{\k}}\label{fgamma_gen} \enspace.
\eeqa
When the source is at a position $\r_s(t)$, its photons reach the decay location along the direction of the vector $\x_{ds} - \r_s(t)$, yielding
\beqa
\theta_s(t) &=& \arctan\left( \frac{\sqrt{(x_d\sin\theta_i - r_{s,x}(t))^2 + r_{s,y}^2(t)}}{x_d\cos\theta_i \pm x_s - r_{s,z}(t)}\right)\nl
\phi_s(t) &=& \arctan\left( \frac{r_{s,y}(t)}{x_d\sin\theta_i - r_{s,x}(t)}\right) \enspace.
\label{thetas_phis_of_t}
\eeqa
The $x$ and $z$ axes are defined based on the position where the source appears to be today as seen from  Earth. We define today to correspond to time $t=0$. The direction of the line of sight is fixed at $\theta_i$, i.e. the angle of the los relative to the $z$-axis, and we also fix $x_d$.
The source has moved in the time it takes light to travel from the source to here.
If today we see the source at a point with coordinates (0, 0, 0), it means that, assuming $\v_s$ to be a constant, it is now at position 
\beq
\r_s(t=0) =(v_{s,x},~v_{s,x},~v_{s,z})\,x_s\enspace.
\eeq
Here $x_s$ is the distance Earth-source when the source was at (0, 0, 0).  In other words, $x_s$ is the distance of the source as it appears today to us and we will assume we know this quantity (eg. from ATNF). 

Then, at a generic time $t$ the source is at a position 
\beq
\r_s(t) =(v_{s,x},~v_{s,x},~v_{s,z})\,(t + x_s )\enspace.
\eeq
Performing the same calculations as Appendix~\ref{sec:signal} using Eq.~\eqref{fgamma_gen}, the axion momentum selected in the decay is
\beqa
p_x &=& -\omega_k \sin\theta_{\k} + (m - \omega_k) \sin\theta_s\cos\phi_s  \nl
p_y &=& (m - \omega_k) \sin\theta_s \sin\phi_s\nl
p_z &=& \omega_k \cos\theta_k + (m - \omega_k) \cos\theta_s
\eeqa

To linear order in $\theta_d$ and $\epsilon$
\beqa
p_x &=& \omega_k\theta_d\cos\phi_s \cos\theta_{\k} - (\omega_k (1-\cos\phi_s) + \epsilon\cos\phi_s)\sin\theta_{k} \nl
p_y &=& (\omega_k\theta_d \cos\theta_{k} + (\omega_k - \epsilon) \sin\theta_{k})\sin\phi_s\nl
p_z &=& \epsilon\ \cos\theta_k + \omega_k \theta_d \ \sin\theta_k \enspace.
\eeqa

Now, what should we use for $t$ in the equations above for a decay happening at a distance $x_d$ along the line of sight?
The quantities $\theta_s(t),~\phi_s(t)$ enter the expression for the echo flux through $f_\gamma$. 
Let's consider a photon emitted by the source at time $t_{em}$ that travels to the decay location and stimulates the decay of an ALP. We then detect one of the photons from this decay today at time $t_{obs}$.
The time of observation must be equal to the time of emission plus the total distance traveled by the photons
\beq
t_{obs} = t_{em} + \tilde x_{ds}(t_{em}) + x_d\enspace.\label{tem}
\eeq
This tells us that for given $t_{obs}$ and $x_d$, we should evaluate Eqs.~\eqref{thetas_phis_of_t}, at $t=t_{em}=t_{obs}-x_d-\tilde x_{ds}(t_{em})$. Eq.~\eqref{tem} can be solved for $t_{em}$ as a function of $x_s$, $x_d$, $\theta_i$ and $\v_s$.

To compute the echo flux  observed from Earth today, we also need expressions for the distance Earth-source and source-decay location 
\beqa
\tilde x_s(t) &=& \sqrt{(v_{s,x}(t+x_s)))^2 + (v_{s,y}(t+x_s))^2 + (\pm x_s - v_{s,z} (t+x_s) )^2 }\nl
\tilde x_{ds}(t) &=& \sqrt{(x_d\sin\theta_i - v_{s,x} (t+x_s) )^2 + (v_{s,y}(t+x_s))^2 + (x_d\cos\theta_i \pm x_s - v_{s,z}(t+x_s))^2}\enspace.\nl
\label{xsxds_of_t}
\eeqa
Notice $x_s = \tilde x_s(-x_s)$ and $x_{ds} = \tilde x_s(-x_s)$, when choosing $t_{obs}=0$.

The echo flux is related to the phase space density of photons at $t_{em}$, at the location where the source was back then, as follows
\beqa
S_\nu(t_{obs}) &\sim& \dot{n}_a(t_{obs}-x_d, \x_{ds})
\sim f_\gamma(t_{obs}-x_d, \tilde x_{ds}(t_{obs}-x_d))\nl
&\sim & 
f_\gamma(t_{em}, \tilde x_{ds}(t_{em}))
\sim f_\gamma(t_{em},r_*)\left(\frac{r_*}{\tilde x_{ds}(t_{em})}\right)^2 \enspace,
\eeqa
where $r_*$ is the radius of the source. Notice that in $f_\gamma(t, x)$, $x$ is the distance from the location where the source was at time $t$.

Finally, we need to relate $f_\gamma(t_{em}, r_*)$ to a known quantity, i.e. source's flux density today on Earth
\beq
S_\nu^{(0)}(t_{obs}) \sim f_\gamma(t_{obs}, \tilde x_s(t_{obs})) = f_\gamma(t_{obs} - \tilde x_s(t_{obs}), r_*)\left(\frac{r_*}{\tilde x_s(t_{obs})} \right)^2\enspace.
\eeq
Then we get
\beq
f_\gamma(t_{em}, r_*)\left(\frac{r_*}{\tilde x_{ds}(t_{em})}\right)^2 
= 
f_\gamma(t_{obs}, x_s(t_{obs}))\,
\frac{f_\gamma(t_{em}, r_*)}{ f_\gamma(t_{obs} - \tilde x_s(t_{obs}), r_*)}
\left(\frac{\tilde x_s(t_{obs})}{\tilde x_{ds}(t_{em})}\right)^2 
\eeq
and we can generalize Eq.~\eqref{flux_density} to
\beq
S_\nu(t) = \frac{\pi g_{a\gamma}^2}{16m}S^{(0)}_\nu(t_{obs}) 
\int d\Omega_i \,B(\Omega_i)\int dx_d\ \rho_a(\x_{ds})\, 
\frac{L_\nu(t_{em})}{L_\nu(t_{obs} - \tilde x_s(t_{obs}))}
\left(\frac{\tilde x_s(t_{obs})}{\tilde x_{ds}(t_{em})}\right)^2 
h(\omega, \x_{ds}, t_{em})\enspace,
\eeq
where $L_\nu(t)$ is the luminosity of the source.

The Doppler shift of the source's radiation as seen from the decay location compared to as seen from Earth is unimportant to the echo if $f_\gamma(\omega_k, x_{ds})$ changes slowly as a function of $\omega_k$.

%%%%%%%%%%%%%%%%%%%%%%%%%%%%%%%%%%%%%%%%%%%%%
%%%%%%%%%%%%%%%%%%%%%%%%%%%%%%%%%%%%%%%%%%%%%
%%%%%%%%%%%%%%%%%%%%%%%%%%%%%%%%%%%%%%%%%%%%%
\section{Implementation}\label{sec:implementation}
To evaluate Eq.~\eqref{flux_density} numerically, we introduce two cartesian coordinate systems: 1) $XYZ$ centered at the Galactic center, the $Z$ axis pointing towards the Galactic north pole (GNP) and the $X$ axis pointing towards the Sun's location; 2) $X'Y'Z'$ obtained by translating $XYZ$ along the $X$ axis by $r_\odot$, where $r_\odot$ is the distance of the Sun from the Galactic center. The system $X'Y'Z'$ is centered at the Sun's location (we neglect the vertical displacement of the Sun from the Galactic plane). A point with Galactic coordinates $(l,b)$ at a distance $d$ from the Sun has cartesian coordinates
\beqa
X' &=& -d \cos b \cos l\nl
Y' &=& -d \cos b \sin l\nl
Z' &=& d \sin b \enspace.
\eeqa
The galactocentric distance of the point $r$ is then given by
\beq
r = \sqrt{X^2 + Y^2 + Z^2} = \sqrt{(X' + r_\odot)^2 + Y'^2 + Z'^2} \enspace.
\eeq
To evaluate Eq.~\eqref{flux_density}, we need to compute $r$ for all points along the line of sight. So, we need to find an equation for the los in the $XYZ$ coordinate system.
Let the source have coordinates $(l_s, b_s)$ and distance from the Sun $x_s$. Then, for the front-light echo, we will center our observations in the direction $(l_s^f, b_s^f) = (l_s, b_s)$, while for the back-light echo we'll have $(l_s^b, b_s^b) = (l_s+\pi, -b_s)$.
The los is displaced on the sky from the direction 
$(l_s^j, b_s^j)$ (with $j=f,b$) by an angle $\theta_i >0$ in some direction $\phi_i$. Let's define $\phi_i = 0$ when the los is displaced with respect to $(l_s^i, b_s^i)$ in the direction of 
growing Galactic latitude and unchanged Galactic longitude, and let $\phi_i$ grow in the clockwise direction when we look at $(l_s^i, b_s^i)$ from the Sun's location. The direction of the los is then $(l_{los}, b_{los}) = (l_s^i - \theta_i \sin\phi_i, b_s^i + \theta_i \cos\phi_i)$.

Let $X_d',\ Y'_d,\ Z'_d$ be the coordinates of a point along the los at a distance $x_s$ from the Sun, then the equations for the los are
\beqa
X' = X_d'\ t\qquad\qquad
Y' = Y_d'\ t\qquad\qquad
Z' = Z_d'\ t\enspace,
\eeqa
and the distance from the Sun  of a point along the los is $x_d = x_s t$ and its galactocentric distance is
\beq
r(t) = \sqrt{x_s^2t^2 + r_\odot^2 + 2X'_d r_\odot t} \enspace.
\eeq

The next step is to calculate the components of the average DM velocity along the axes of the system $xyz$ defined in Figures~\ref{fig:geom1} and~\ref{fig:geom2}. To do so we express the unit vectors $\hat{x},\ \hat{y},\ \hat{z}$ in the $X'Y'Z'$ system.
They are given by
\beqa
\hat{x} &=& (\sin b_s \cos l_s \cos \phi_i \pm \sin l_s \sin \phi_i,
\ \sin b_s \sin l_s \cos \phi_i \mp \cos l_s \sin\phi_i,
\ \cos b_s \cos \phi_i)\nl
\hat{y} &=& (-\sin b_s \cos l_s \sin \phi_i \pm \sin l_s \cos \phi_i,
\ -\sin b_s \sin l_s \sin \phi_i \mp \cos l_s \cos\phi_i,
\ -\cos b_s \sin \phi_i)  \nl
\hat{z} &=& \mp(-\cos b_s \cos l_s,\ -\cos b_s \sin l_s,\ \sin b_s )\enspace.
\eeqa
From these, we can calculate the components of the average ALP velocity in the $xyz$ system and then obtain $\langle v_\parallel\rangle$  and $\langle v_\perp \rangle$ through the rotation matrix Eq.~\eqref{rotation_xz}.

\section{Collinear emission}\label{sec:collin}
Starting from the Boltzmann equation for the collinear emission Eq.~\eqref{boltzmann_collinear}
\beqa
\frac{d^3\dot{n}_a}{dq^3}
 &=& -
\frac{1}{(2\pi)^3}\frac{1}{2\omega_q}
 \int \frac{d^3p}{(2\pi)^3}\frac{f_a(\p\,)}{2m_a}\ 
 \frac{d^3k_1}{(2\pi)^3}\frac{1}{2\omega_{k}}
 (2\pi)^4\delta^{(4)}(k + q - p) \sum_{\lambda}|\mathcal{M}_0|^2
 (1+f_\gamma(\q, \lambda)) 
 \enspace,\nl
\eeqa
we proceed to integrate over the momentum of the echo photon traveling back toward the source
\beqa
\frac{d^3\dot{n}_a}{dq^3}
 &=& -
\frac{1}{(2\pi)^3}\frac{1}{(2\omega_q)^2}
 \int \frac{d^3p}{(2\pi)^3}\frac{f_a(\p\,)}{2m_a}\ 
 (2\pi)\delta(2\omega_q - m - p \cos\theta_p) \sum_{\lambda}|\mathcal{M}_0|^2
 (1+f_\gamma(\q, \lambda)) 
 \enspace.\nl
\eeqa
We have chosen the direction of $\q$ to be that of the positive $z$-axis and as usual we keep corrections of order $p$ only when they are relevant to the frequency or direction of the photons.
Plugging in the expression for the matrix element
\beqa
\frac{d^3\dot{n}_a}{dq^3}
 &=& -
\frac{1}{(2\pi)^2} \frac{g_{a\gamma}^2m_a}{4} f_\gamma(\q, \lambda)
 \int \frac{d^3p}{(2\pi)^3}f_a(\p\,)\ 
\delta(\epsilon - p \cos\theta_p) 
 \enspace.
\eeqa
Let's focus on the integral over the axion momentum $I = \int \frac{d^3p}{(2\pi)^3}f_a(\p\,)\ \delta(\epsilon - p \cos\theta_p)$
\beqa
I
&=&\frac{n_a}{(2\pi)^{3/2}\delta p^3}\int dp\,p^2\ e^{-\frac{p^2 + \langle p\,\rangle^2}{2\delta p^2}}
\int d\Omega_p\delta(\epsilon- p \cos\theta_p)e^{\frac{p\langle p\,\rangle}{\delta p^2}\cos\theta_{\p\langle\p\,\rangle}}
 \enspace,
\eeqa
where $\cos\theta_{\p\langle\p\,\rangle} = \cos\theta_p\cos\theta_{\langle p \rangle} +\sin\theta_p\sin\theta_{\langle p \rangle}\cos\phi_p$.

We can integrate over the azimuthal angle
\beqa
\int d\phi_p\ e^{\frac{p\langle p\,\rangle}{\delta p^2}\sin\theta_p\sin\theta_{\langle p \rangle}\cos\phi_p}
=2\pi I_0\left(\frac{p\langle p\,\rangle}{\delta p^2}\sin\theta_p\sin\theta_{\langle p \rangle}\right)\enspace,
\eeqa
where $I_0$ is the modified Bessel function of the first kind of order 0. 
Let' define the component of the axion momentum parallel and perpendicular to the los: $\langle p_\parallel\rangle = \langle p\,\rangle \cos\theta_{\langle p \rangle}$ and $\langle p_\perp\rangle = \langle p\,\rangle \sin\theta_{\langle p \rangle}$, and take the integral over $\cos\theta_p$ using the delta-function
\beqa
\int d\Omega_p\delta(\epsilon- p \cos\theta_p)e^{\frac{p\langle p\,\rangle}{\delta p^2}\cos\theta_{\p\langle\p\,\rangle}} 
&=&
\frac{2\pi}{p}\int d\cos\theta_p\  \delta\left(\cos\theta_p - \frac{\epsilon}{p}\right)
e^{\frac{p\langle p_\parallel\rangle}{\delta p^2} \cos\theta_p} 
 I_0\left(\frac{p\langle p_\perp\rangle}{\delta p^2}\sin\theta_p\right)\nl
&=&
\frac{2\pi}{p}
e^{\frac{\epsilon\langle p_\parallel\rangle}{\delta p^2} } 
 I_0\left(\frac{\langle p_\perp\rangle}{\delta p^2}\sqrt{p^2-\epsilon^2}\right)
\enspace,
\eeqa
Then 
\beqa
I &=&
\frac{2\pi n_a}{(2\pi)^{3/2}\delta p^3}\ e^{-\frac{\langle p\,\rangle^2}{2\delta p^2}}\ e^{\frac{\epsilon\langle p_\parallel\rangle}{\delta p^2} } 
\int dp\,p\ e^{-\frac{p^2}{2\delta p^2}}
 I_0\left(\frac{\langle p_\perp\rangle}{\delta p^2}\sqrt{p^2-\epsilon^2}\right)\nl
% &=&
% \frac{2\pi n_a}{(2\pi)^{3/2}\delta p^3}\ e^{-\frac{\langle p\,\rangle^2}{2\delta p^2}}\ e^{\frac{\epsilon\langle p_\parallel\rangle}{\delta p^2} } 
% \delta p^2 \ e^{-\frac{\epsilon^2}{2\delta p^2}}\ e^{\frac{p_\perp^2}{2\delta p^2}}\nl
% &=&
% \frac{n_a}{\sqrt{2\pi}\delta p}\  e^{\frac{\epsilon\langle p_\parallel\rangle}{\delta p^2} }  \ e^{-\frac{\epsilon^2}{2\delta p^2}}\ e^{\frac{\langle p_\parallel\rangle^2}{2\delta p^2}}
% \nl
&=&
\frac{n_a}{\sqrt{2\pi}\delta p}\  e^{-\frac{(\epsilon - \langle p_\parallel\rangle)^2}{2\delta p^2}}
\enspace.
\eeqa
As expected, the transverse average axion momentum has no effect, and the decay rate is the largest for $\epsilon = \langle p_\parallel\rangle$.

We have now our desired result
% \beqa
% \frac{d^3\dot{n}_a}{dk^3}
%  &=& -
% \frac{1}{(2\pi)^2} \frac{g_{a\gamma}^2m_a}{4} f_\gamma(\k, \lambda)
% \frac{n_a}{\sqrt{2\pi}\delta p}\  e^{-\frac{(\epsilon - \langle p_\parallel\rangle)^2}{2\delta p^2}}
%  \enspace.
% \eeqa
% and thus
\beqa
\frac{d\dot{n}_a}{d\omega_q d\Omega_q}
 &=& -
n_a \frac{g_{a\gamma}^2m_a^3}{64\pi^2} f_\gamma(\q, \lambda)
\frac{1}{\sqrt{2\pi}\delta p}\  e^{-\frac{(\epsilon - \langle p_\parallel\rangle)^2}{2\delta p^2}}
 \enspace.
\eeqa

\end{document}